\documentclass[useAMS,usenatbib]{mn2e}
\usepackage{graphicx}
\usepackage{epsfig}

\newcommand{\HII}{H\,{\sc ii}}

\newcommand{\OIII}{[O\,{\sc iii}]}

\newcommand{\NII}{N\,{\sc ii}}
\def\p0{\phantom{0}}

\title[A New Population of Planetary Nebulae Discovered in the Large Magellanic Cloud (I)]
{A New Population of Planetary Nebulae Discovered in the Large
Magellanic Cloud (I): Preliminary Sample}
\author[Warren. A. Reid and Quentin A. Parker]{Warren. A. Reid$^{1}$\thanks{e-Mail:
warren@ics.mq.edu.au; tosame@bigpond.net.au } and Quentin A.
Parker $^{1,}$$^{2}$\thanks{e-Mail: qap@ics.mq.edu.au}\\
$^{1}$Department of Physics, Macquarie University, Sydney, NSW 2109, Australia\\
$^{2}$Anglo-Australian Observatory, PO Box 296, Epping, NSW 1710
Australia}
\begin{document}

\date{Accepted 2004. Received 2004 ; in original form 2004}

\pagerange{\pageref{firstpage}--\pageref{lastpage}} \pubyear{}

\maketitle

\label{firstpage}

\begin{abstract}
We report our initial discovery of 73 new planetary nebulae (PNe)
in the Large Magellanic Cloud (LMC) following confirmatory 2dF
spectroscopy on the Anglo-Australian Telescope (AAT). Preliminary
candidate sources come from a 10 per cent sub-area of our new
deep, high resolution H$\alpha$ map of the central 25$^{\circ}$
square of the LMC obtained with the UK Schmidt Telescope (UKST).
The depth of the high resolution map was extended to
$R_{equiv}\sim22$ for H$\alpha$
($4.5\times10^{-17}ergs~cm^{-2}~s^{-1}~$\AA$^{-1}$) by a process
of multi-exposure median co-addition of a dozen 2-hour H$\alpha$
exposures. The resulting map is at least 1-magnitude deeper than
the best wide-field narrow-band LMC images currently available.
This depth, combined with our selection technique, has also led to
the discovery of extended AGB halos around many new and previously
known LMC PNe for the first time. Once complete, our new survey is
expected to triple the LMC PN population and have significant
implications for the LMC PN luminosity function, kinematics,
abundance gradients chemical evolution and, via study of the AGB
halos, the initial to final mass relation for low to intermediate
mass stars.
\end{abstract}


\section{Introduction}

The study of planetary nebulae (PNe), including determination of
their physical properties and luminosity function, has been a
difficult task in our galaxy due to inherent problems of accurate
distance determination and the biases introduced by variable
interstellar absorption. Such problems can be alleviated by
studying PNe in a nearby galaxy such as the Large Magellanic Cloud
(LMC).

The LMC, at a distance of 50kpc, is essentially a thin
($\sim$500pc) disk inclined at only 35~degrees to our line of
sight (van der Marel \& Cioni 2001) so all LMC PNe can be
considered to reside at a similar distance. Knowledge of this one
fundamental property permits direct estimation and comparison of a
significant number of LMC PN physical parameters. It also affords
considerable astrophysical benefits by providing a single
environment in which PN mass loss history (and hence that of
intermediate to low mass stars) can be studied in detail in the
context of both stellar and galactic evolution (Jacoby, 2005).
Distance estimates are being further refined thanks to a variety
of observational relations for stars in our own galaxy which are
being applied to those in the LMC (e.g. via RR Lyrae variables,
eclipsing binary stars, main sequence fitting and Mira variables -
see Madore \& Freedman 1998 for a review) promising further
precision of measured PN physical parameters.

The observed reddening and extinction towards the LMC is
comparatively low and uniform (e.g. Kaler \& Jacoby, 1990) so that
a more complete PN population can, in principle, be obtained. The
LMC galaxy is also sufficiently small in angular extent that it
can be studied in its entirety with a wide-field telescope such as
the UK Schmidt Telescope (UKST). Furthermore, it is sufficiently
close to enable PNe to be detected and resolved using current
ground-based telescopes. LMC PNe can be observed with HST allowing
morphological studies (Shaw et al. 2001) and central star
measurements (e.g. Villaver et al. 2003). This data can be
compared with sophisticated model calculations to allow refinement
of nebular photo-ionization codes. These in turn allow the most
accurate determination of absolute fluxes, measurement of chemical
abundance variations (e.g. Dopita \& Meatheringham 1991),
determination of the mass and age of the central stars and the
mass, physical size and age of the nebulae with their associated
extended halos. Such precise calculations and the ability to
directly compare nebula age to that of the central star will allow
tests of models of late stellar evolution that would not otherwise
be possible (Jacoby 2005).

The Magellanic Clouds are rich with compact emission-line sources
such as WR, Be, LBV, Of, T Tauri and symbiotic stars, VV Cephei
systems and PN. Prior to our recent work however, the number of
LMC PNe remained modest ($\sim$300) and comprised a small fraction
of the expected total ($1000\pm250$, Jacoby 1980). Furthermore,
the current samples comprise a heterogeneous compilation from
surveys with varying depth, selection technique, detection
efficiency and spatial coverage. This makes them of limited use
for the unbiased estimation of key PN parameters necessary to
study their evolution, that of their host galaxy and the
associated inter-stellar medium enrichment and mass loss history
of their stars in a quantitative way.

However, based on the preliminary success reported here, a near
complete sample of the LMC PN population is promised such that
truly meaningful quantitative determinations of the PN luminosity
function, distribution, abundances, kinematics and crucially
mass-loss history, can be estimated precisely for the first time.






\section{Previous LMC PN surveys: a brief review}

Most early work on the LMC PN population was undertaken with a
variety of modest, wide-field telescopes fitted with
objective-prism dispersers. The first study was that of Henize
(1956) who published the positions of 415 nebulae including 97
point-like emission sources which were considered to be PN
candidates. Lindsay (1963) published 109 point-like H$\alpha$
emission sources with no continuum from a sample of about 1000
emission objects. Where emission lines other than H$\alpha$ were
seen, the objects were classified as PN (65 objects). Westerlund
and Smith (1964) surveyed 100 square degrees of the LMC with a
blue objective prism on a 20/26 inch Schmidt telescope to produce
a catalogue of 42 LMC PNe with photometric data, approximate
coordinates and manually drawn finding charts. As a result of
their efforts, the more luminous PNe in both clouds were
identified and subjected to detailed studies such as those by
Feast (1968) and Webster (1969). To this point however, the nature
of most of the fainter objects detected in the above-mentioned
surveys remained unknown or ambiguous.

Sanduleak, MacConnell and Phillip (1978), after a decade of
amassing a collection of various types of deep objective-prism
plates, produced another catalogue (SMP) comprising 102 PNe
covering a large fraction of the LMC. The coordinates of the SMP
survey are approximate to $\sim$1 arcmin with no finding charts.
Although positions had to be re-established, it was the most
complete list of confirmed PNe at that time and was therefore
preferred to previous surveys.

Jacoby (1980) directly imaged 4 central regions of the LMC in
\OIII~ and H$\alpha$ with the Cerro Tololo 4m telescope. This was
deeper than any previous survey and produced 41 faint PN
candidates. Equatorial co-ordinates and finding charts were
presented with uncertainties $<$0.6~arcsec. Monochromatic fluxes
were derived and used to determine the luminosity function for LMC
PN as faint as 6 magnitudes below the brightest known. Although 2
candidates were previously known, a follow-up survey by Boroson \&
Liebert (1989) confirmed 13 as PNe.


Morgan (1984) used a new 2$^\circ$11$^{\prime}$ objective-prism on
the 1.2/1.8m UKST to study spectra of most of the known PNe in
both clouds. This prism gave a dispersion of 880{\AA} mm$^{-1}$ at
H$\beta$ 4861{\AA} allowing good separation of the \OIII~ doublet
lines from each other and from H$\beta$ under good seeing
conditions. A number of faint new PNe were discovered in regions
not surveyed by Jacoby (1980). Where possible, the spectra were
assigned excitation classes following the classification of Feast
(1968).

Sanduleak (1984) found an additional 25 PN candidates (commonly
denoted Sa) using deeper objective-prism plates centered on the
central region of the LMC. Five objects were later rejected. These
were double checked by Leisy et.al. (1997) who found that
candidate Sa103 is a galaxy (as suspected by Sanduleak 1984) with
z=0.035 while four objects could not be found again. 


Morgan \& Good (1992) and Morgan (1994), published 86 and 54 new
PN candidates respectively from UKST objective prism plates to a
continuum magnitude of B$_{\textit{j}}$ $\sim$19.5. Due to the
increased depth, continua could now be seen on both new and many
of the previously known PNe. Some PNe were known to have diameters
large enough to be resolvable on UKST plates (Jacoby 1980, Wood et
al. 1987). Approximately 12 per cent of PNe appeared extended on
their plates, with full width diameters of $\geq$4 arcsec.
Coordinate errors are quoted as typically $\pm$2-3 arcsec.


\section{Problems with existing LMC emission object catalogues and the need for a new survey}

The history of LMC PN searches has been biased through strong
selection effects as well as being affected by identification and
astrometric problems. This demonstrates the need for a homogenous,
deep, well constructed LMC PN census with high astrometric
integrity. After Henize (1956), subsequent H$\alpha$ surveys were
progressively deeper through use of better instrumental
resolutions and observational configurations. While this improved
sensitivity resulted in an increased number of emission-line
detections, surveys continued to use single objective-prism plates
and separated emission-line stars from nebulae according to
whether a continuum could be seen adjacent to the H$\alpha$ line.
The deeper surveys moreover do not cover the full $\sim$25$^\circ$
square field occupied by the main bar of the LMC. Although the
UKST is an excellent instrument for obtaining a complete survey of
the LMC because of its wide-field coverage, the best dispersion
available at H$\alpha$ with an objective-prism is 2050{\AA}/mm
which is poor compared to normal spectroscopy.

H$\alpha$ emission alone cannot distinguish between object types.
Furthermore, many of the emission-line objects in these Schmidt
surveys have not been observed spectroscopically (Morgan 1998). A
single wide-field direct exposure in H$\alpha$ easily reveals
extended objects such as \HII~ regions. At the distance of the
LMC, however, most PNe are effectively point sources thereby
making them indistinguishable from normal stars. What was required
was a high resolution LMC map in the H$\alpha$ line and
neighbouring continuum so that the emission line objects could be
highlighted, as the difference between the two images. The
continuum could be either a separate direct image from a broad
waveband which may include but dilute H$\alpha$ emission, or a
single exposure through an objective prism. The last technique has
been the most commonly employed. In crowded fields such as the
LMC, however, it has suffered from the overlapping of sources and
modest depth. Most survey work has therefore been done through
restricted wavebands sufficiently narrow to minimise this problem.
The faint stars which merged into the sky background on blended
objective prism plates, then became separate individual objects.
Where images continued to be blended, complex software was
required to de-blend and match plates.

Due to the time lag often present between exposures of the same
region, variable stars were revealed, appearing similar to PN
candidates. They were identified firstly by comparison with other
plates and also by comparison with other emission lines such as
\OIII~5007\AA ~for PNe and HeII~4686\AA ~for WR and symbiotic
stars. As an example, several symbiotic stars discovered on the
basis of objective-prism spectra with HeII~4686 emission and a red
stellar continuum were found to be members of the above mentioned
H$\alpha$ catalogues (Morgan 1992).

To date $\sim$300 LMC PNe are known and have published
\OIII~5007\AA ~photometry. Leisy et al. (1997) have used the SERC
J sky Atlas to provide $\textit{Bj}$ broadband magnitudes for all
the LMC PNe but these objects are extremely faint in the broad
$\textit{Bj}$ survey waveband ($\sim$21), with respect to the
neighbouring population. In relative terms they are many times
brighter and easier to measure in H$\alpha$ light.

Until now there was no complete, deep, H$\alpha$-based PN
catalogue for the LMC. However, the overall brightness of these
objects in the light of H$\alpha$ makes this wavelength an
efficient discovery tool (Morgan 1998). Deep searches began in
1993 involving H$\alpha$ filters with CCD detectors on large
telescopes (Azzopardi 1993). The drawback however is that these
systems only covered very small areas and were designed to produce
highly detailed surveys of emission-line stars and small PNe in
selected areas around objects such as clusters and associations. A
deep, high resolution H$\alpha$ survey of the whole LMC has been
needed to identify the many emission-line stars fainter than the
14.5 mag. limit of the Bohannan \& Epps (1974) survey, to detect
low excitation PN not seen in the published \OIII~5007\AA ~based
surveys and extend the PN searches to fainter limits.

Another ongoing LMC emission-line survey is being conducted by
Smith (1998) which uses several interference filters including
H$\alpha$ in conjunction with a CCD detector on the Curtis Schmidt
Telescope. The survey covers most of the central parts of the
cloud, but, at 3 arcsec resolution, is designed for extended
objects and is not ideal for point sources.

\begin{figure}
  \includegraphics[width=.45\textwidth]{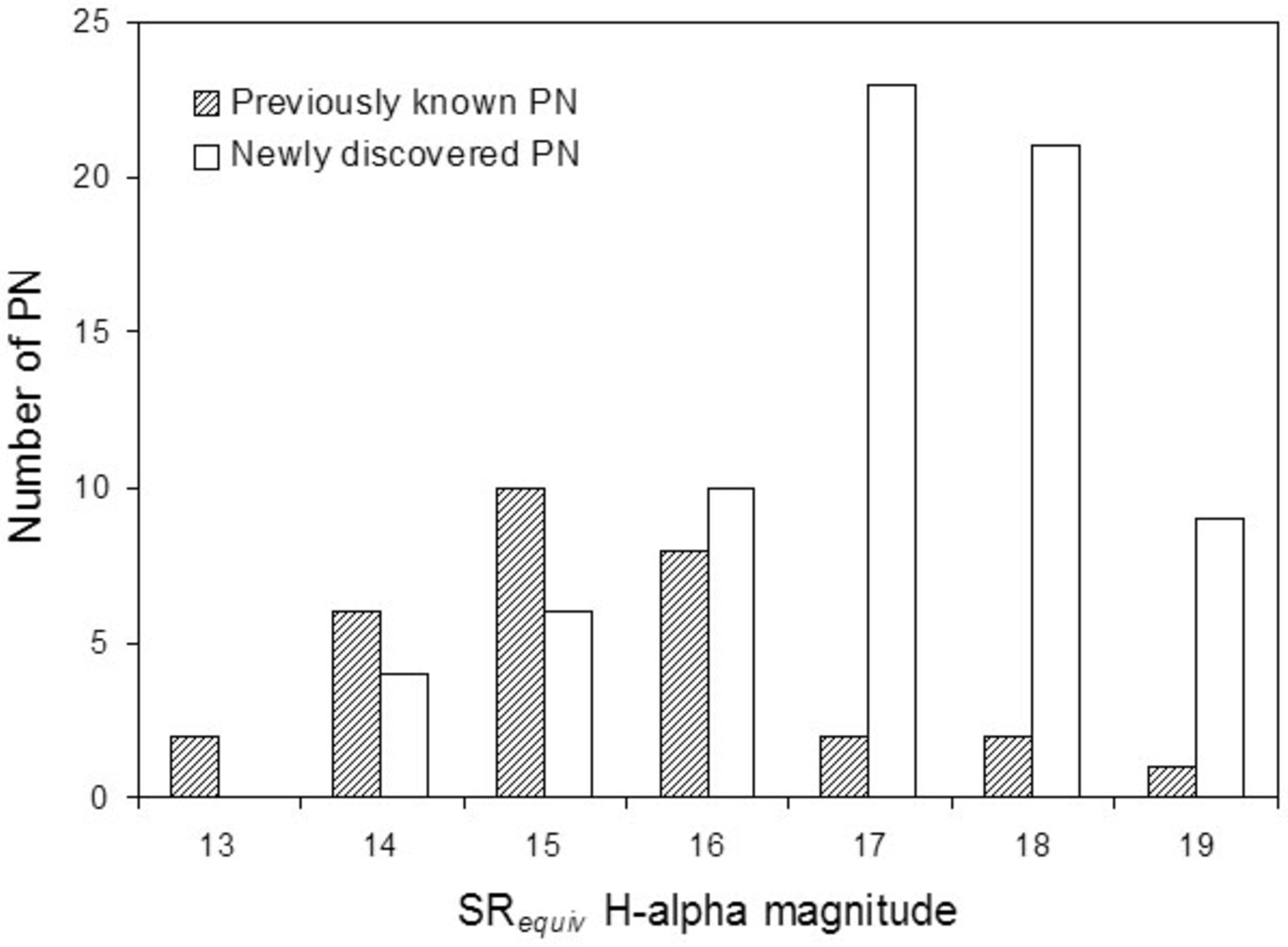}\\
  \caption{Previously known versus. newly confirmed PNe by magnitude in the preliminary $\sim$3.25${^\circ}$ square region observed. An overall increase in number and depth by 3 magnitudes is evident.}
  \label{Figure 1}

  \includegraphics[width=.47\textwidth]{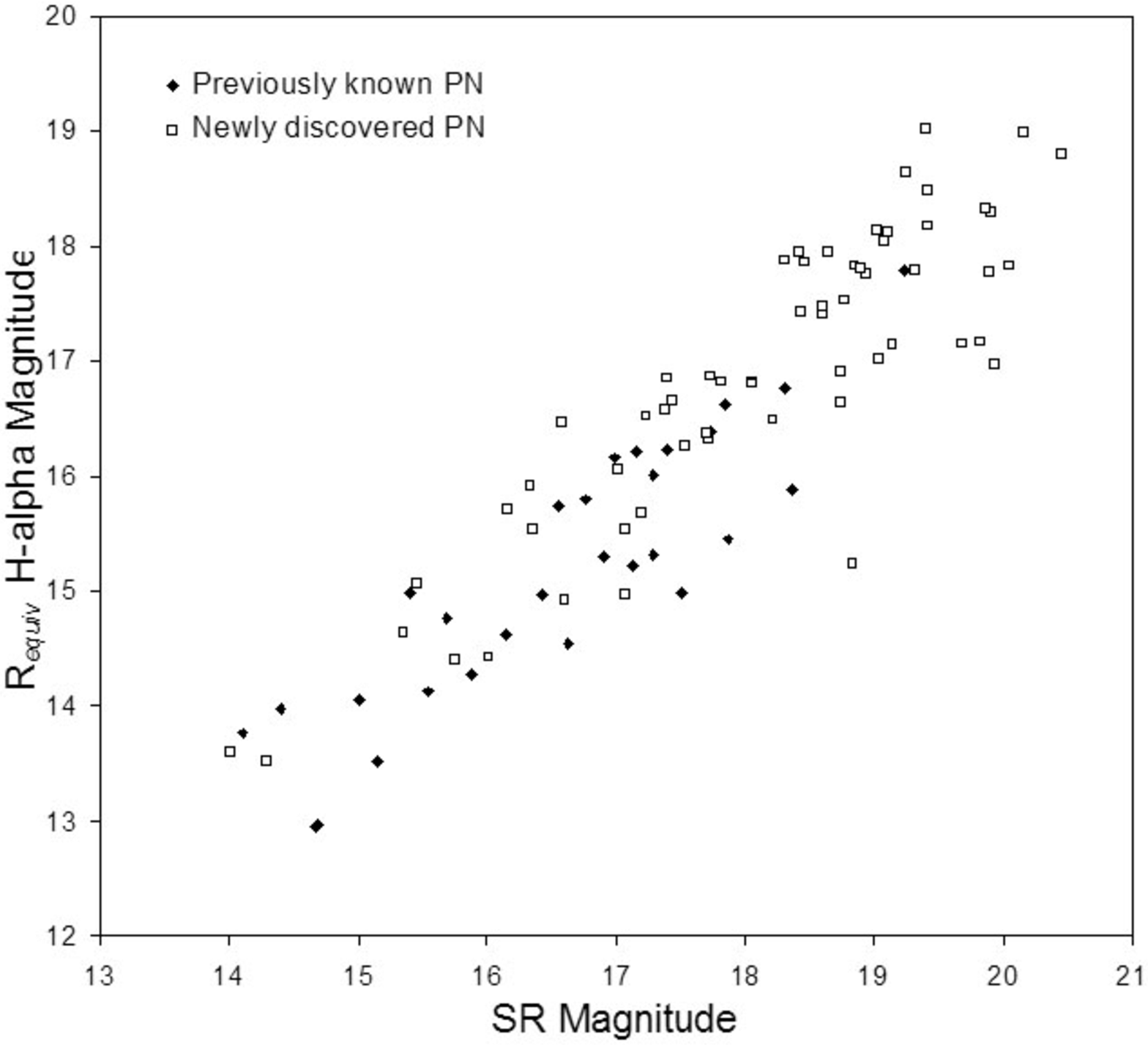}\\
  \caption{Plot of SR versus R$_{equiv}$H$\alpha$ magnitude for previously known and newly confirmed PNe in the current observed region. Previously known PNe show a slightly wider magnitude differential and occupy the bright end of the plot. New PNe extend the plot to fainter limits.}
  \label{Figure 10}
  \end{figure}

\subsection{The new AAO/UKST H$\alpha$ map of the LMC}

As part of the AAO/UKST H$\alpha$ (+ effectively [NII]) survey of
the Southern Galactic Plane (Parker et al. 2005) an equivalent 40
field mini-survey of the entire LMC, SMC and surrounding regions
was undertaken. The survey used an exceptional quality,
monolithic, 70\AA ~FWHM H$\alpha$ interference filter (Parker \&
Bland-Hawthorn 1998) and fine grained Tech-Pan film as detector
(Parker and Malin 1999) to yield a survey with a powerful
combination of sensitivity, resolution and area coverage. The fast
f/2.48 converging beam of the UKST is effectively even faster
off-axis. Consequently, the pass-band of the narrow-band H-alpha
filter shifts slightly to the blue at the field edges relative to
the centers of the surveyed fields. The impact of this effect,
though, is small, affecting only the wings of the [NII] 6584\AA
~line (see Parker et al. 2005 for details).

Over the last few years, we have specially constructed additional
deep, homogeneous, narrow-band H$\alpha$ and matching broad-band
`SR' (Short Red) maps of the entire central 25$^\circ$ square of
the LMC. These unique maps were obtained from co-adding twelve
well-matched UK Schmidt Telescope 2-hour H$\alpha$ exposures and
six 15-minute equivalent SR-band exposures on the same field using
high resolution Tech-Pan film. The `SuperCOSMOS' plate-measuring
machine at the Royal Observatory Edinburgh (Hambly et al. 2001)
has scanned, co-added and pixel matched these exposures creating
10$\mu$m (0.67~arcsec) pixel data which goes 1.35 and 1 magnitudes
deeper than individual exposures, achieving the full canonical
Poissonian depth gain, e.g. Bland-Hawthorn, Shopbell \& Malin
(1993). This gives a depth $\sim$21.5 for the SR images and
$R_{equiv}\sim$22 for H$\alpha$
($4.5\times10^{-17}ergs~cm^{-2}~s^{-1}~$\AA$^{-1}$) which is at
least 1-magnitude deeper than the best wide-field narrow-band LMC
images currently available. The influence of variable stars is
alleviated and emulsion defects removed by median stacking
exposures taken over a 3 year period. An accurate world
co-ordinate system was applied to yield sub-arcsec astrometry,
essential for success of the spectroscopic follow-up observations.

The survey is now photometrically calibrated (see section
~\ref{section 6}), and can be used to provide R and
R$_{\textit{equiv}}$ H$\alpha$ magnitude estimates for any
emission sources detected. Table~\ref{Table 1} gives the details
of all the UKST exposures used for stacking in this survey. The
co-ordinates of the film centre in the B1950 equinox are RA
$05^h22^m0~$ and~ DEC $-69^{\circ}00'$. Exposure times and dates
of exposures are also given.


\begin{figure}
\begin{flushleft}
  \includegraphics[width=.47\textwidth]{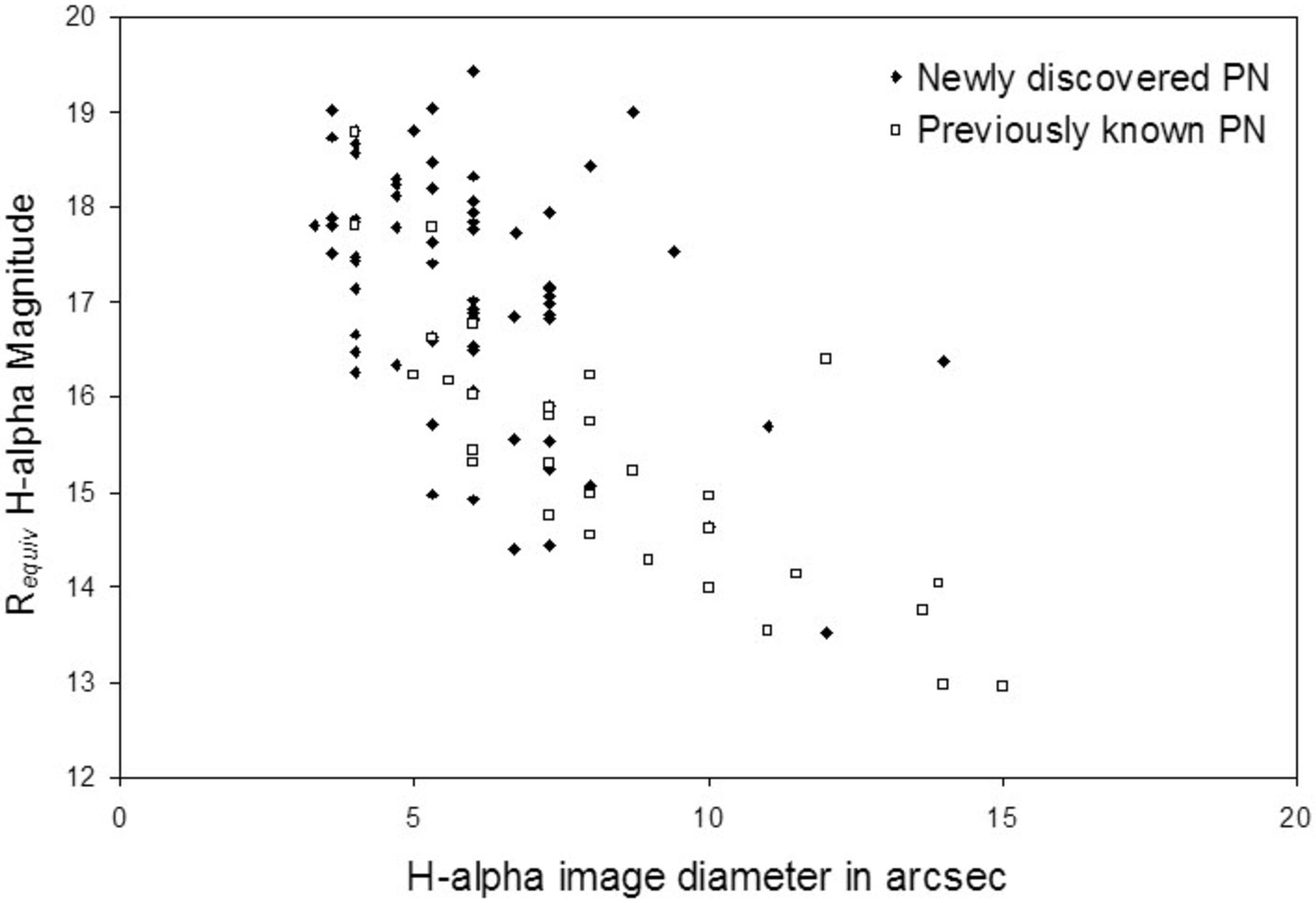}\\
  \caption{R$_{equiv}$H$\alpha$ magnitude versus image derived diameter for both previously known and newly discovered PNe in the preliminary $\sim$3.25${^\circ}$ square region observed. This plot reveals the large number of faint (low surface brightness) and highly evolved PNe in the new sample. The brighter of the previously known PNe with true diameters $\leq$3 arcsec are unresolvable by the UKST and incur a growth in image diameter as a function of luminosity (see section \ref{section 4.1}).}\label{Figure 12}
  \end{flushleft}
\begin{flushleft}
  \includegraphics[width=.47\textwidth]{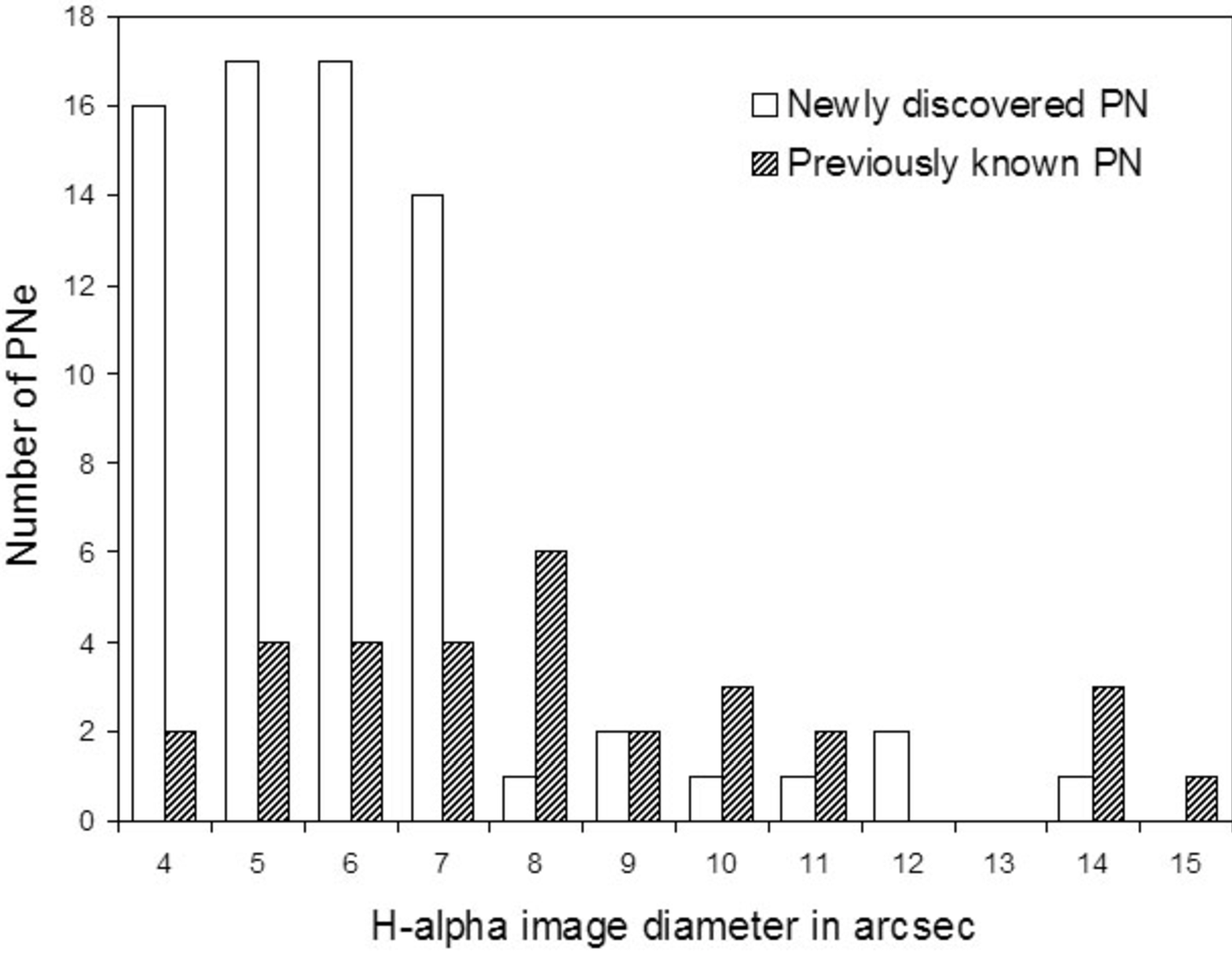}\\
  \caption{H$\alpha$ image diameter versus number for both previously known and newly discovered PNe in the preliminary $\sim$3.25${^\circ}$ square region observed revealing the large number of new PNe with resolvable diameters between 4 and 7 arcsec. Several of the previously known low surface brightness PNe also have resolvable diameters within this range (see section \ref{section 4.1}).}\label{Figure 20}
  \end{flushleft}
  \end{figure}


\begin{table}
\caption{Details of the Tech-Pan H$\alpha$ and SR exposures taken
on field MC22 at RA $05^h22^m0~$ and~ DEC $-69^{\circ}00'$ (B1950)
for this deep stacking project.}
\begin{tabular}[]{|c|c|c|c|c|c|}
  \hline
  Plate No. &  Date & LST  & Filter & Exp. time & Grade \\
  \hline
  OR17777  F   & 971021 & 0320  & OG 590 & 150 & a  \\
  HA17778  F   & 971021 & 0349  & HA659 & 1200 & aI  \\
  HA17787  F   & 971023 & 0355  & HA659 & 1200 & a  \\
  OR17790  F  & 971024 & 0341  & OG 590 & 150 & a  \\
  HA17791  F  & 971024 & 0408  & HA659 & 1200 & a  \\
  OR17793  F  & 971025 & 0516  & OG 590 & 150 & a  \\
  OR17814  F  & 971122 & 0516  & OG 590 & 150 & a  \\
  HA17815  F  & 971122 & 0545  & HA659 & 1200 & a  \\
  HA17840  F  & 971223 & 0404  & HA659 & 1200 & aI  \\
  HA17852  F  & 971228 & 0440  & HA659 & 1200 & a  \\
  OR17853  F  & 971228 & 0654  & OG 590 & 150 & a  \\
  HA18160  F  & 981113 & 0511  & HA659 & 1200 & a  \\
  HA18179  F  & 981129 & 0345  & HA659 &  600 & a  \\
  OR18180  F &  981129 & 0501  & OG 590 & 150 & a  \\
  HA18181  F  & 981129 & 0530  & HA659 & 1200 & a  \\
  HA18678  F  & 991207 & 0513  & HA659 & 1200 & a  \\
  HA18709  F  & 000105 & 0426  & HA659 & 1200 & a  \\
  HA18712  F  & 000110 & 0454  & HA659 & 1200 & a  \\
  \hline
  \label{Table 1}
\end{tabular}
\end{table}

\begin{figure*}
  \includegraphics[width=.42\textwidth]{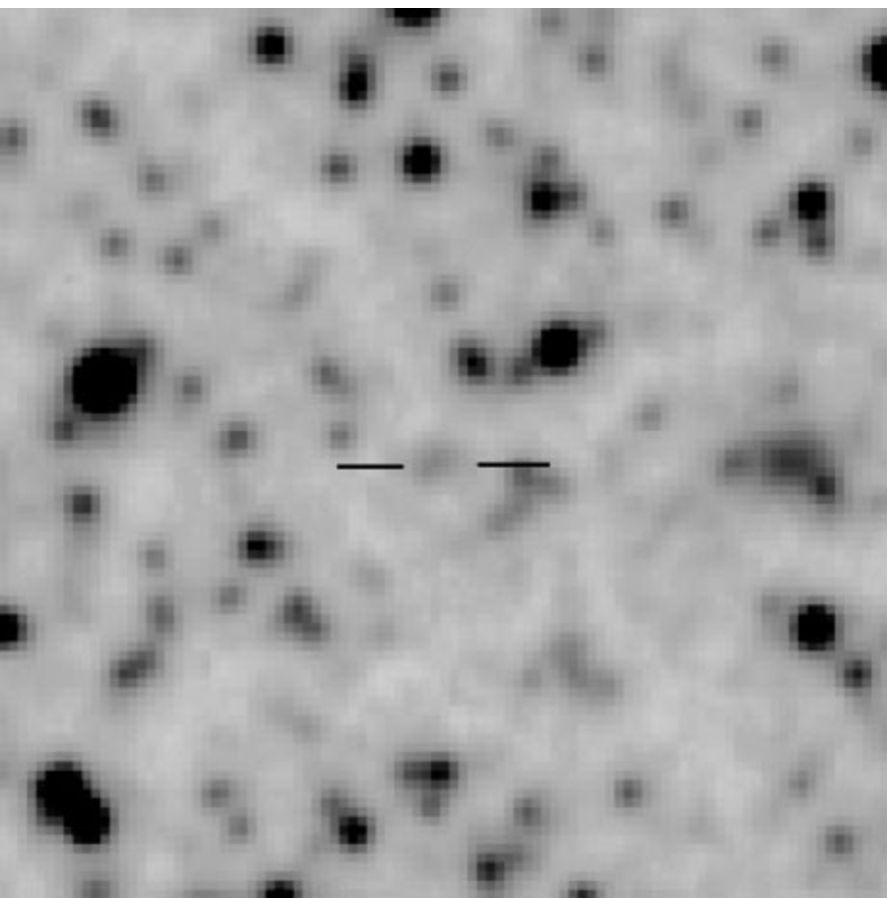}
  \includegraphics[width=.43\textwidth]{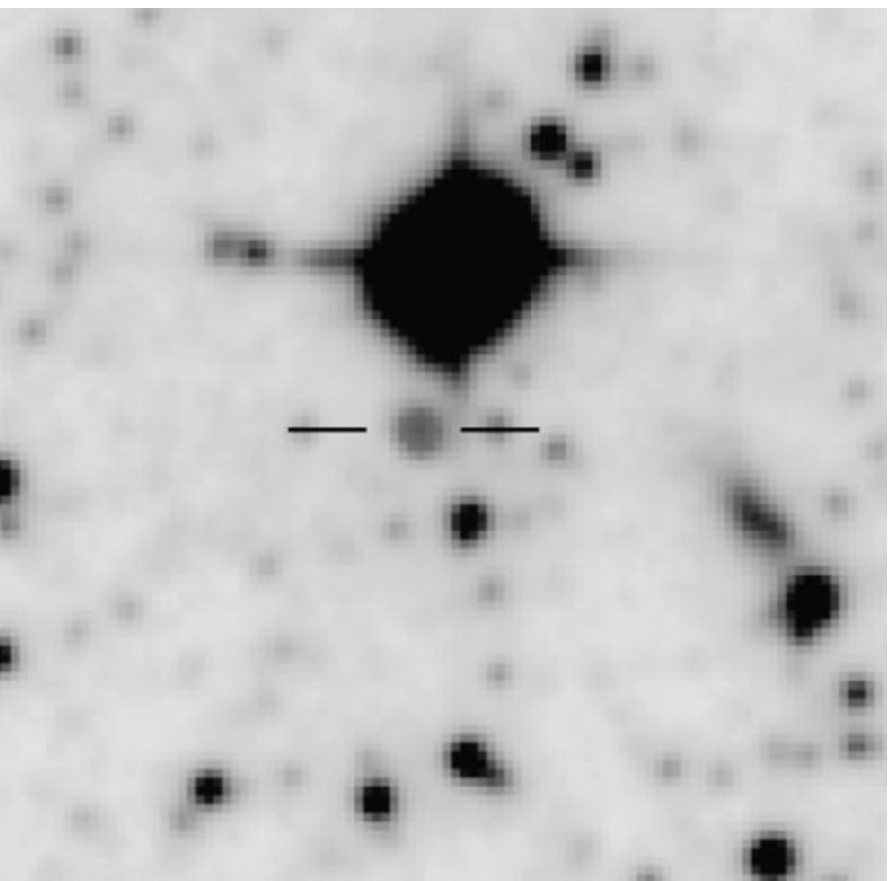}
  \caption{AAO/UKST stacked H$\alpha$/SR combined 50 $\times$ 50 arcsec image of RP442 (left) and RP530 (right).}
  \label{Figure 5}

  \includegraphics[width=.46\textwidth]{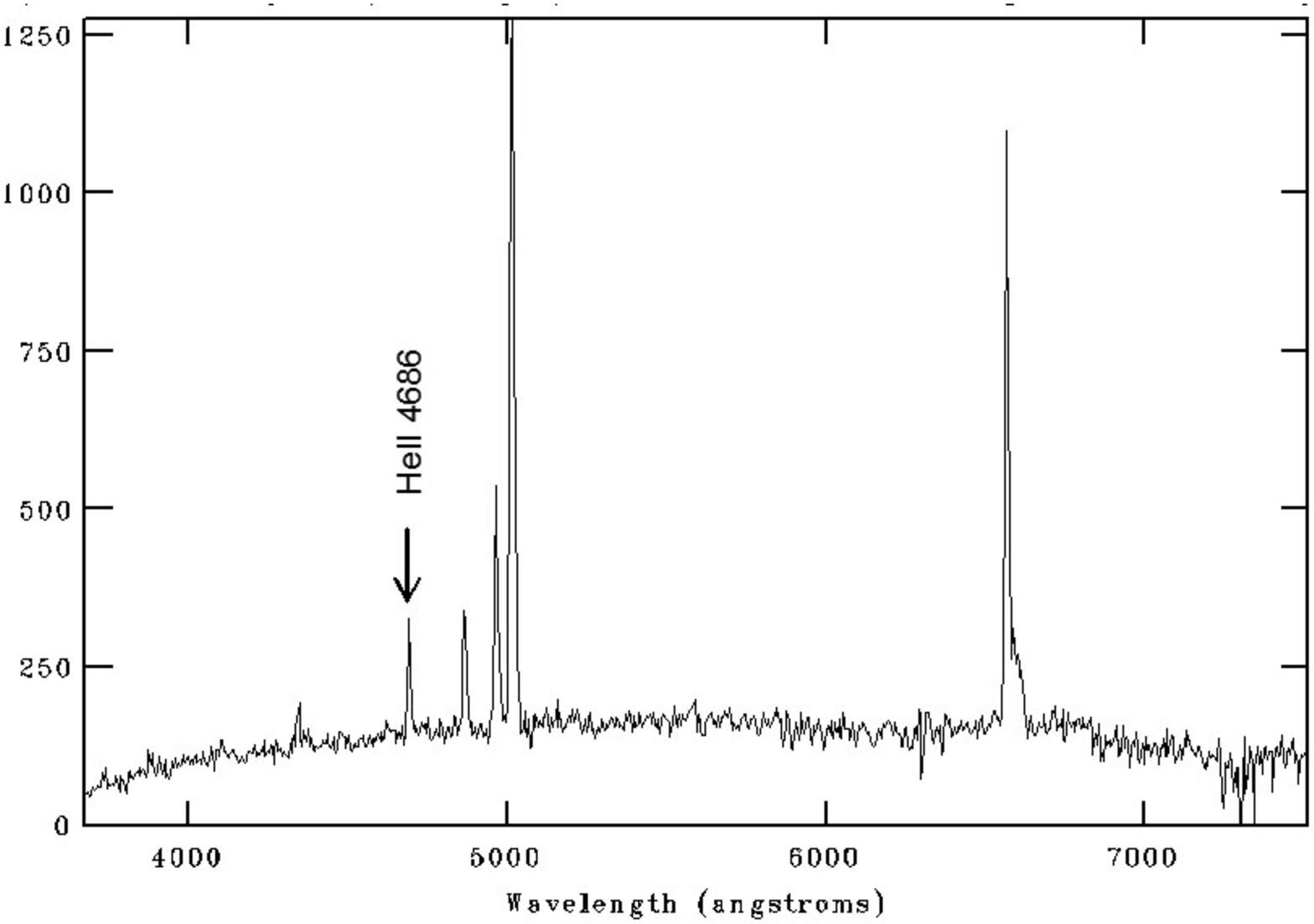}
  \includegraphics[width=.47\textwidth]{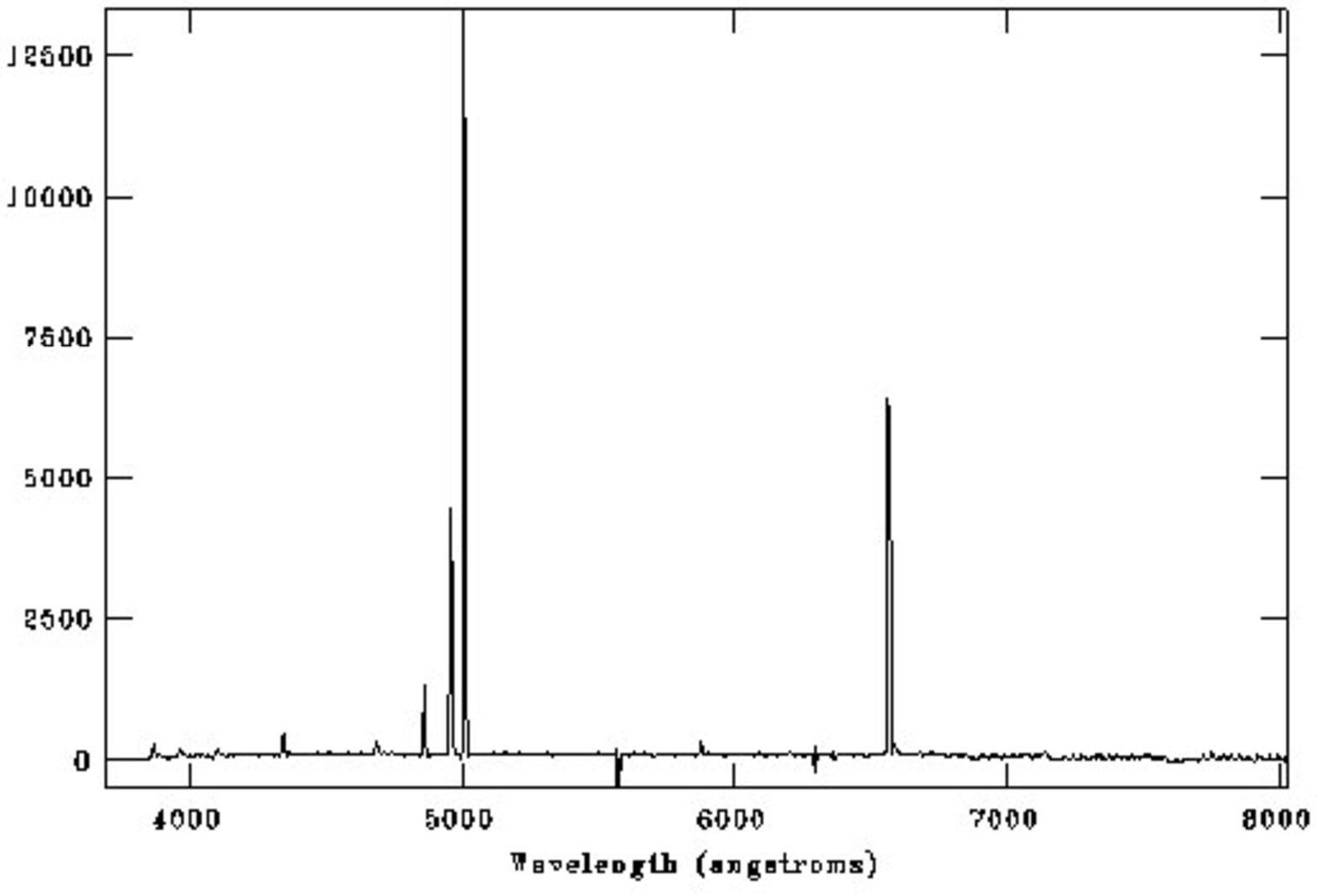}
  \caption{Low-resolution spectra of the newly confirmed PN, RP442 (left) and RP530
(right) of Fig.~\ref{Figure 5}, showing a high
H$\alpha$/[\NII]6583 ratio.} \label{Figure 7}
\end{figure*}

\begin{figure*}
  \includegraphics[width=.435\textwidth]{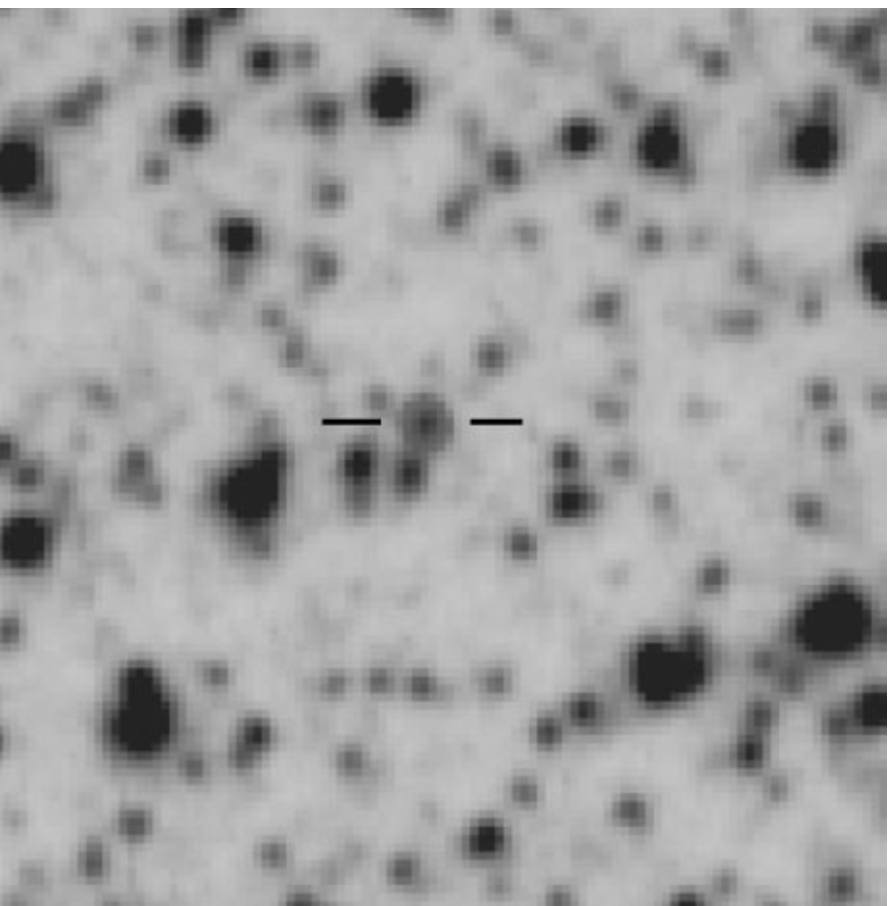}
  \includegraphics[width=.44\textwidth]{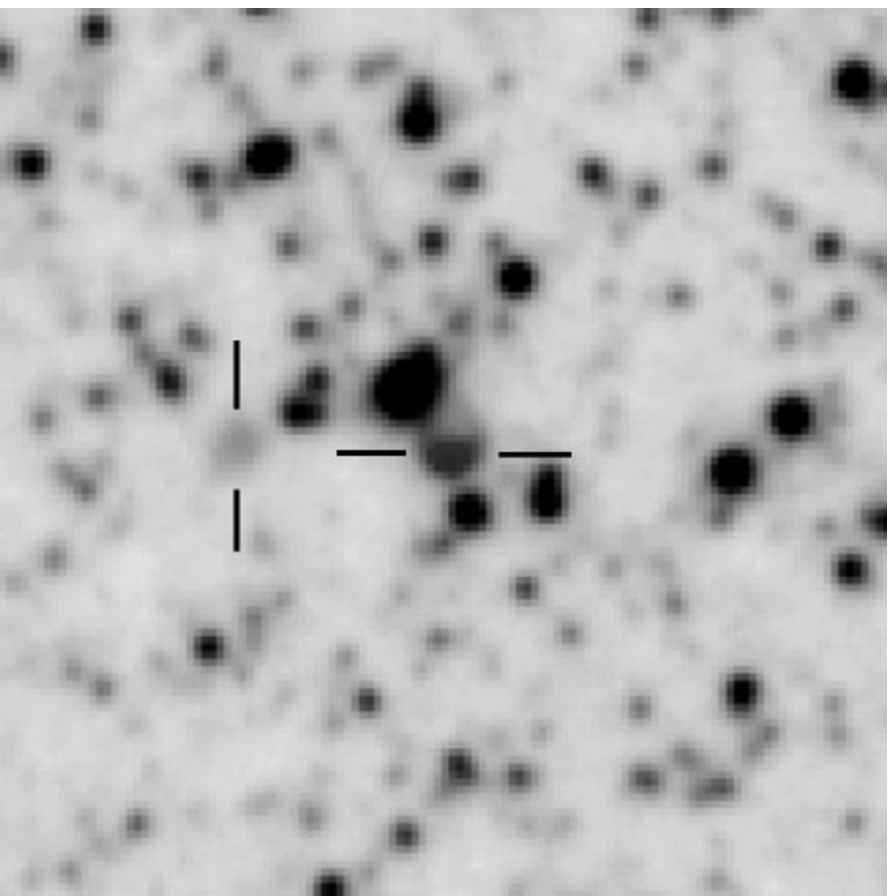} \caption{AAO/UKST stacked
H$\alpha$/SR combined 50 $\times$ 50 arcsec image of RP93 (left)
and RP143 (right), which also includes RP142 to the left of
centre.} \label{Figure 6}

  \includegraphics[width=.485\textwidth]{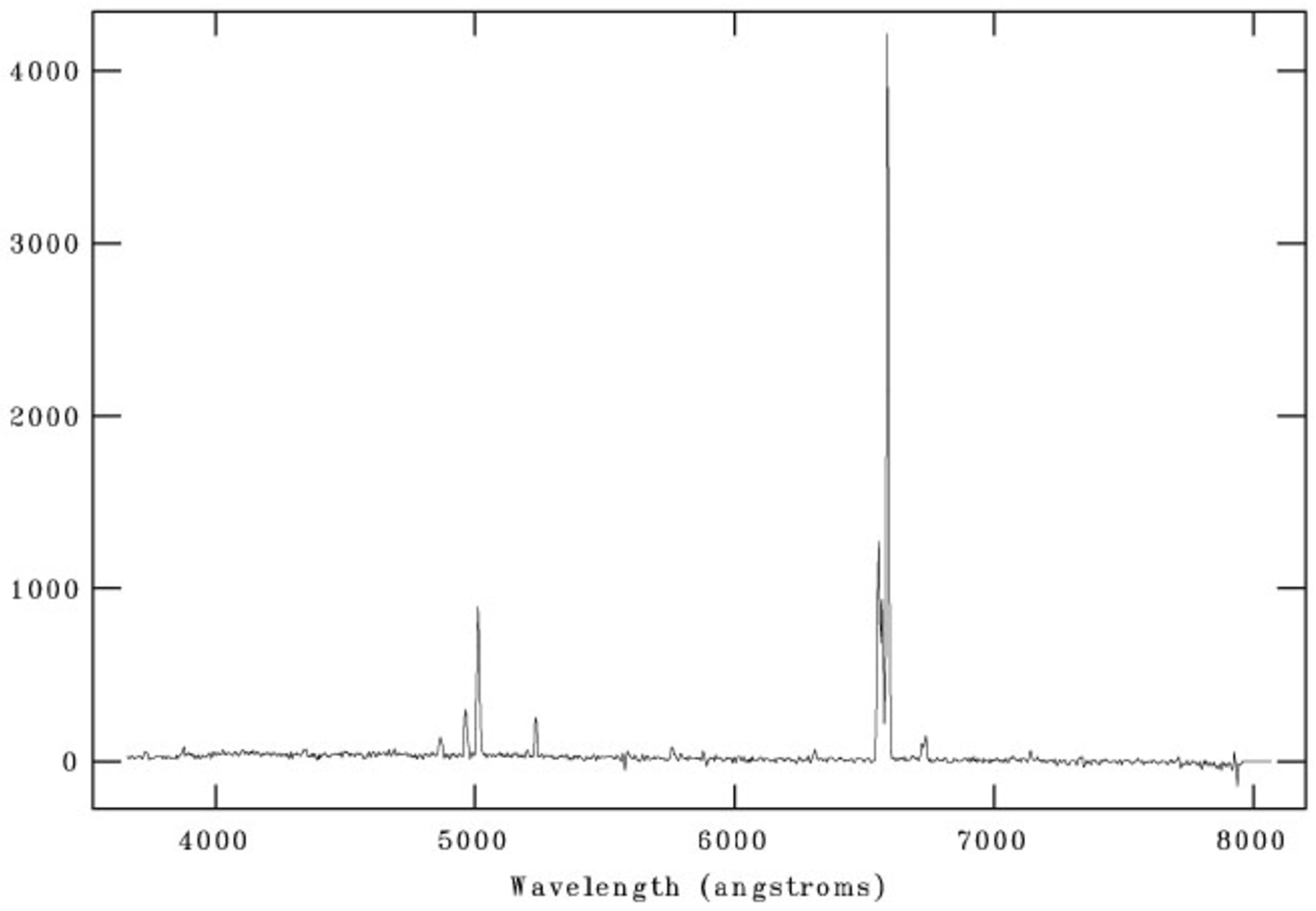}
  \includegraphics[width=.485\textwidth]{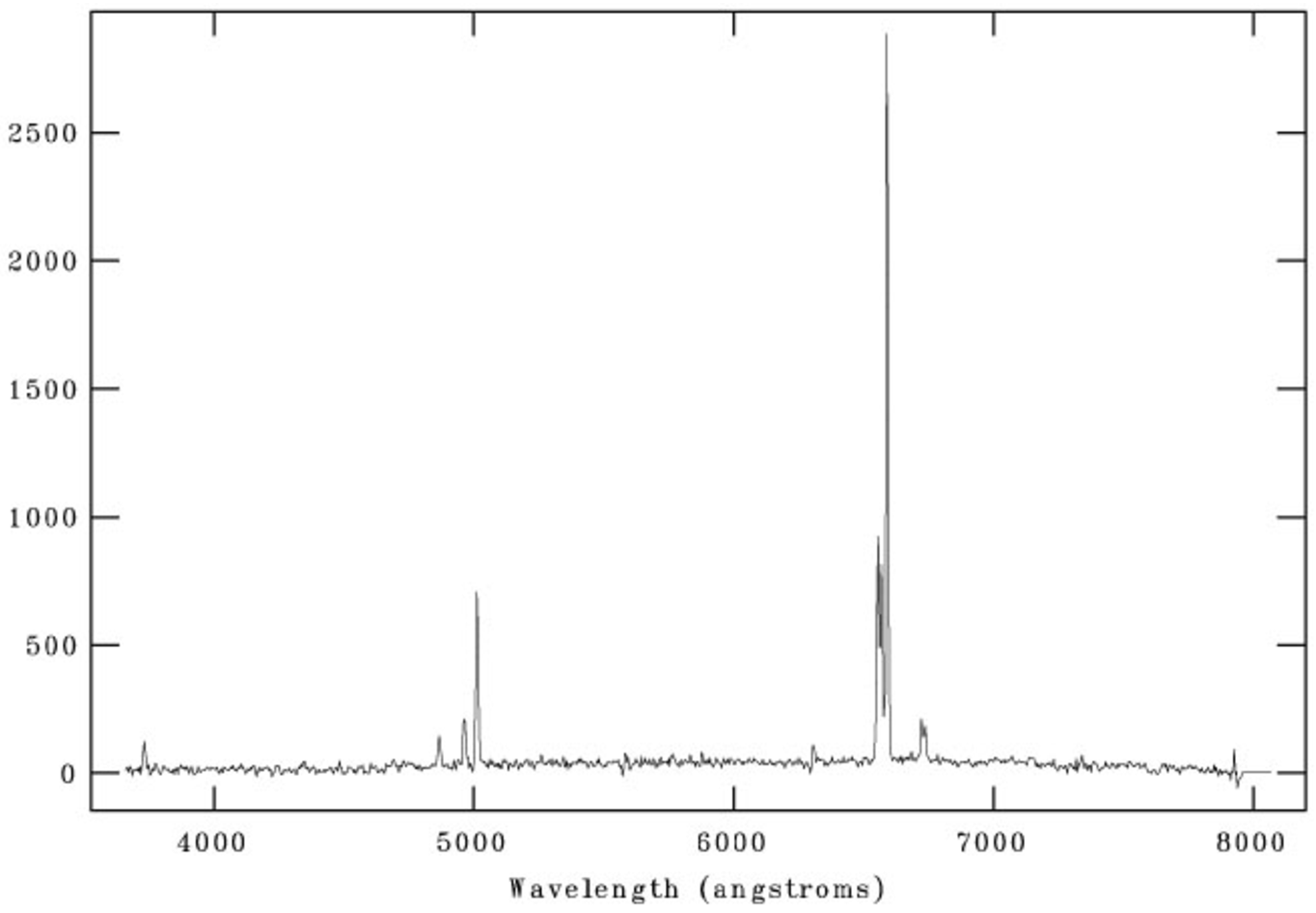}
  \caption{Low-resolution spectra of the newly confirmed PN, RP93 (left) and RP143 (right)
of Fig.~\ref{Figure 6}, showing a high [\NII]6583/H$\alpha$
ratio.} \label{Figure 9}
\end{figure*}

\section{PN candidate selection technique}

The full 25 square degree area of our LMC deep map was subdivided
into 16 separate, non overlapping image cells on a 4 x 4 grid,
each with $\textit{x}$ and $\textit{y}$ dimensions of
approximately 1$^{\circ}$18$^{\prime}$. Each image cell was stored
as a Flexible Image Transport System ({\small FITS}) file,
comprising 6985 x 6985 pixels. The total pixel image size for the
full 25 square degrees of the LMC is 27940 x 27940 pixels
requiring 2.9GB of disk space.

A world co-ordinate system was applied to the images through the
K-ords program within the {\small KARMA} software package (Gooch
1996). A co-ordinate solution was achieved by matching small point
sources at approximately 15 positions across a downloaded,
astronomically calibrated SuperCOSMOS sub-image to the same
positions on the SR and H$\alpha$ UKST stacked images. An
algorithm within the program then applied exactly the same
co-ordinates to the new images. In terms of digitization,
H$\alpha$/SR stacked images were perfectly matched to the
broad-band images available on line through the SuperCOSMOS web
site (http://www.wfau.roe.ac.uk/sss/). The transfer of
co-ordinates was achieved with an accuracy of 0.087 arcsec (0.13
pixels) at the centre of the images, degrading to a displacement
of 1.34 arcsec ($<2$ pixels) at the edges. Every source was
subsequently double-checked and corrected for position with
reference to the same area in the standard online R-band
SuperCOSMOS digital catalogue.

\begin{figure*}
  \includegraphics[width=.444\textwidth]{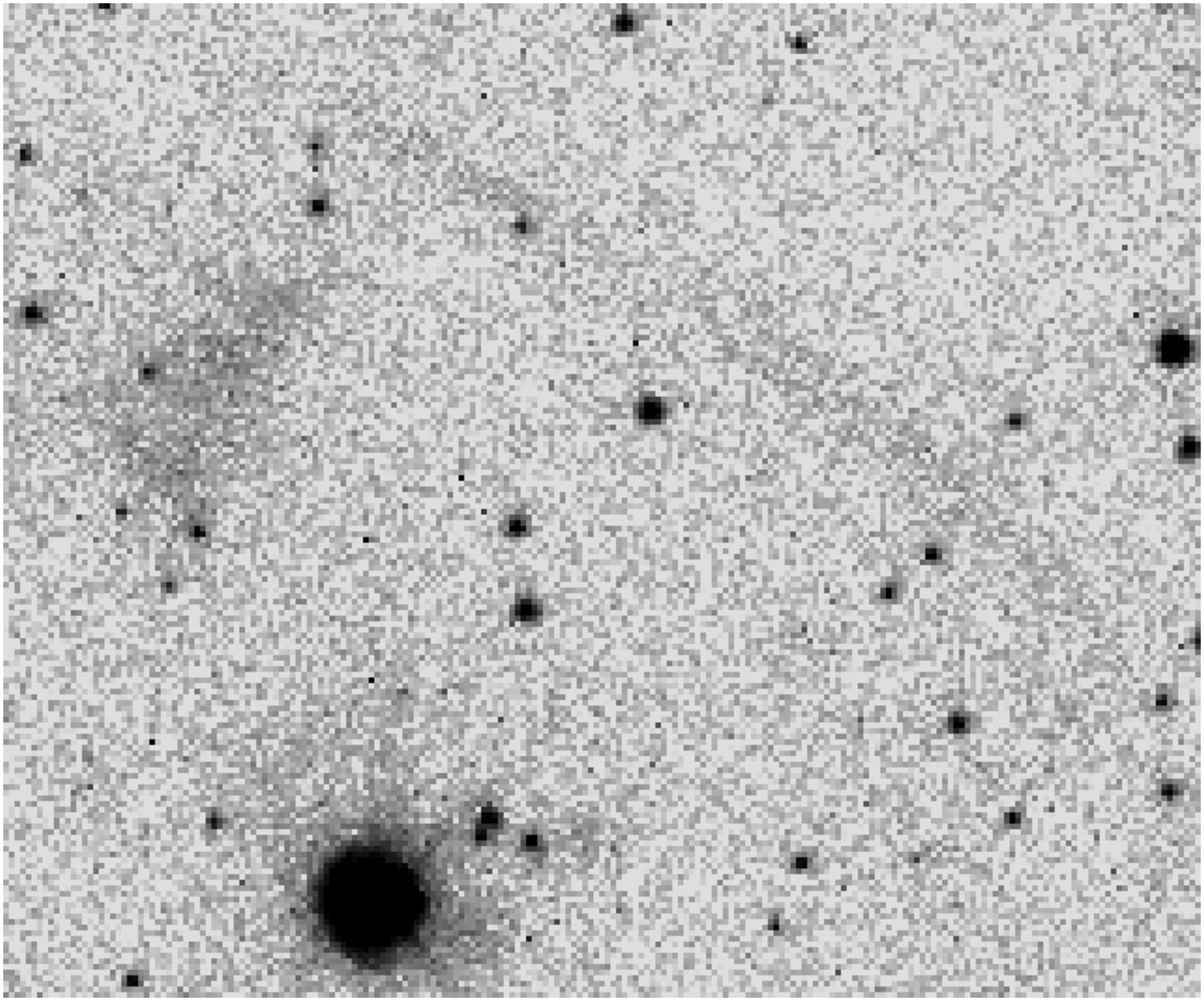}
  \includegraphics[width=.44\textwidth]{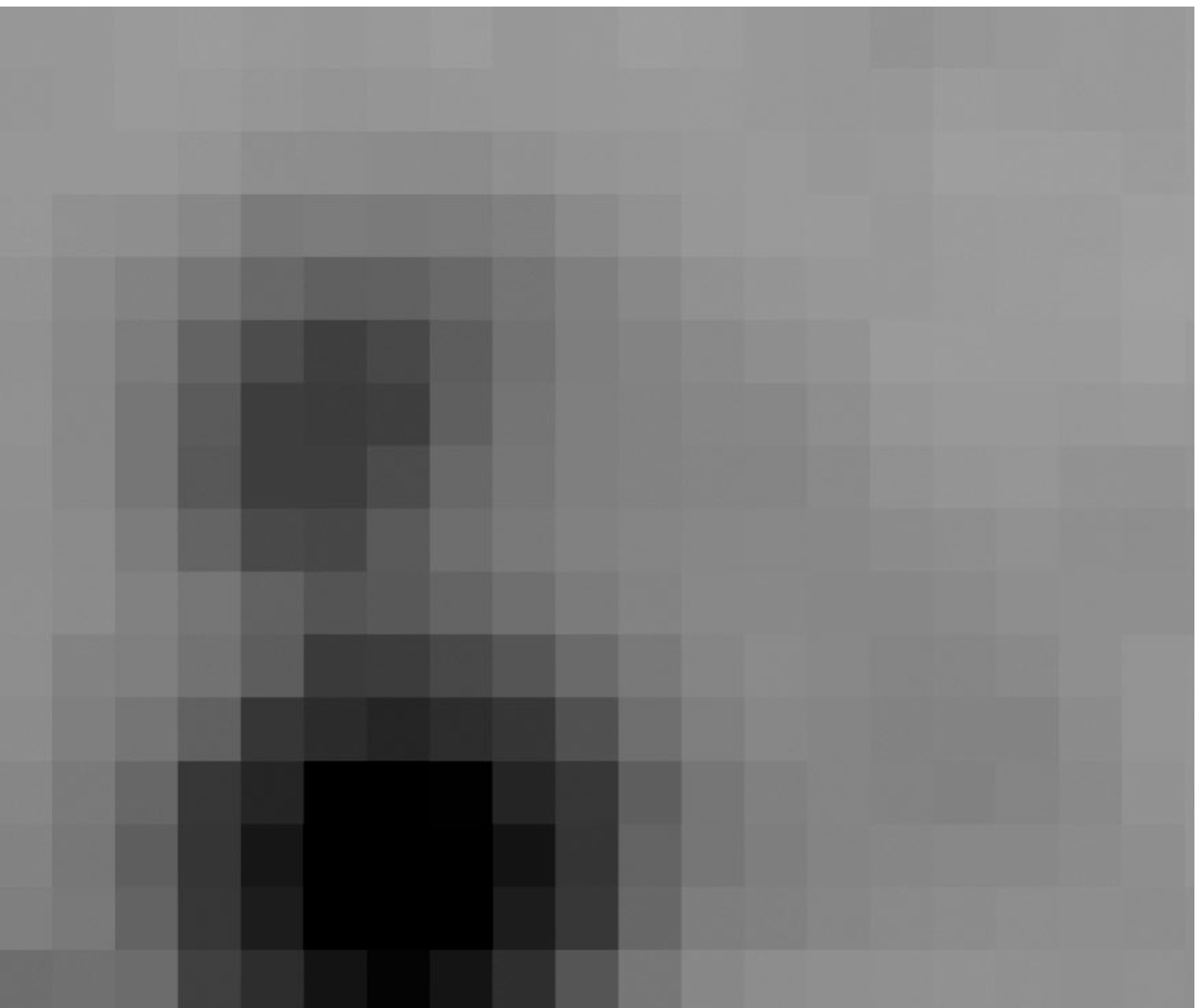}
\caption{Left: HST H$\alpha~9 \times 12$ arcsec image of LMC PN
SMP27 and its associated halo. The PN is the bright blob at lower
left and the halo is the detached faint arc structure starting at
the top left. Right: Matching deep AAO/UKST combined H$\alpha$/SR
stacked image exposure showing the same low-surface brightness arc
structure above the PN.} \label{Figure 3}

  \includegraphics[width=.44\textwidth]{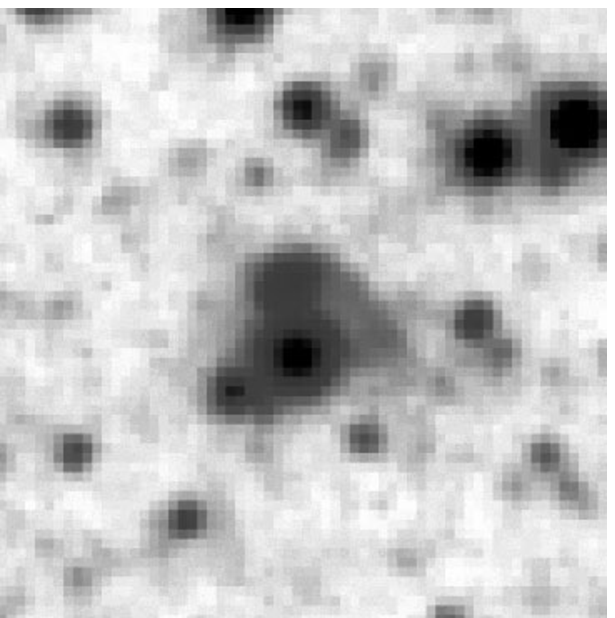}
  \includegraphics[width=.44\textwidth]{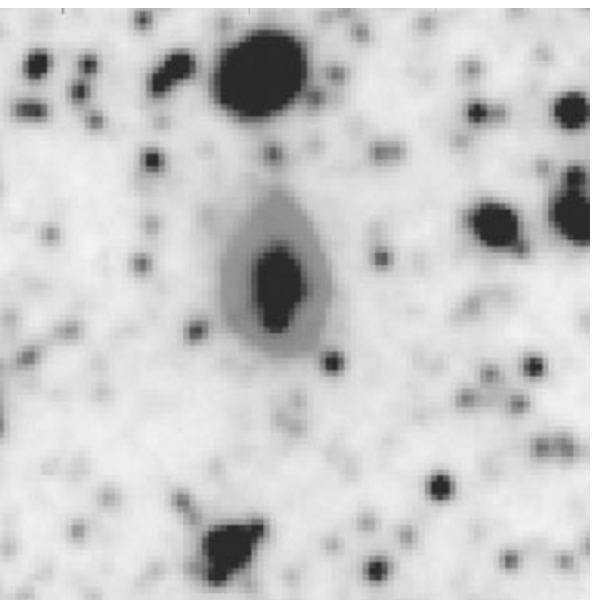}
  \caption{Left: Our wider $30\times30$ arcsec deep stacked H$\alpha$/SR combined image of LMC PN
SMP27 revealing the halo's full extent and surrounding
environment. Right: RP105; a typical example of an AGB halo for a
new LMC PN. The size is $50\times50$ arcsec which is small
compared to the 109,000 square arc-minute coverage for the full
map. See http://www.physics.mq.edu/$\sim$qap/halos.htm for further
examples.} \label{Figure 4}
\end{figure*}

Candidate emission sources were found using an adaptation of a
technique available within {\small KARMA}. The SR images were
coloured red and merged with the H$\alpha$ narrow-band images
coloured blue. Careful selection of software parameters allowed
the intensity of the matched H$\alpha$ and SR {\small FITS} images
to be perfectly balanced allowing only peculiarities of one or
other pass-band to be observed and measured. Normal continuum
stars become a uniform rounded pink/purple colour. Emission
objects such as \HII~regions and the halos surrounding PNe are
strongly coloured blue while the broader point spread function
(PSF) of emission-line stars in the light of H$\alpha$ compared to
their SR counterparts, allows them to be easily detected by the
narrow extent of their faint blue auras, made clearly visible by
the colour-merging software. For examples of how PN are revealed,
see Figs.~\ref{Figure 5},~\ref{Figure 6} \&~\ref{Figure 4}.


Upon detection of a new source in the H$\alpha$/SR stacked and
merged images, we checked and recorded the position in J2000
co-ordinates and the diameter observed on the stacked image in
arcsec. We then commented whether or not the object shows a
central source in the SR, considered the general appearance in
relation to verified objects and added comments. All candidates
were checked in SIMBAD and against all existing PN catalogues for
any previous detections. Previously known objects were included in
the overall catalogue which now represents a complete description
of all the discrete, $<20$$^{\prime\prime}$ diameter emission
objects seen in the UKST deep stacked H$\alpha$ map. Using the
above criteria, we began by broadly rating objects as either a
candidate PN, emission-line star, variable star, \HII~region,
emission object or late-type star with a designation of either
known, probable, possible or unlikely for each category. This
probability rating was assigned to each newly detected emission
candidate prior to spectroscopic confirmation to aid target
prioritization.

The veracity of the technique has been powerfully demonstrated
(see Fig.~\ref{Figure 1}) firstly, by the ease of independent
re-identification of all previously known PNe in each studied
region, no matter how faint (e.g. PNJ07 at m$_{B}$$\sim$21.7) and
secondly, by the success of our AAT 2dF pilot study which
confirmed $\sim$73 new LMC PNe (tables~\ref{Table 2}
and~\ref{Table 4} and section ~\ref{section 7}).

While we are extending the faint limit of the LMC PN luminosity
function (Fig.~\ref{Figure 1}), we are also contributing to the
bright end with several confirmed new PNe brighter than R$\sim$15.
The SR versus SR$_{equiv}$H$\alpha$ magnitude graph
(Fig.~\ref{Figure 10}) also reveals the depth of the new sample
when H$\alpha$/SR magnitudes are compared. While most of the
previously known PNe are scattered at the brighter (lower left)
side of the plot, many are considerably brighter in H$\alpha$ than
SR as expected.

\subsection{PN diameters on the stacked images}

Fig.~\ref{Figure 12} combines the previously known and newly
confirmed PNe in the studied region to show their comparative
magnitude to size relation and growth function as measured in the
stacked images. Diameters for both the previously known and newly
discovered PNe were individually measured on the SR and H$\alpha$
digitized maps using the Starlink {\small GAIA} package. All
measurements were made at the default logarithmic intensity value
in each band. While the UKST is able to resolve any unsaturated
and extended object $>$4 arcsec diameter, point source objects
bright in H$\alpha$ increase in size on the H$\alpha$ map as a
function of flux and exposure time. With the addition of
photographic film as the recording media, the bright point sources
have an image diameter which grows above that expected from a pure
point source (including a seeing disk convolution) as a direct
function of magnitude. A PSF increase by bright PNe will therefore
comprise a portion of the extended emission halos detected
surrounding PNe in H$\alpha$. The magnitudes of previously known
PNe display a tight relationship to the H$\alpha$ diameter
indicating how luminosity is affecting image size at the bright
end of the luminosity function. The newly confirmed PNe, on the
other hand, display a wider, more extended and fainter range of
magnitudes per measured image diameter. Since they are unsaturated
low surface brightness nebulae and halos, they are a truer
reflection of angular size. They are therefore extending the
previous limits to fainter and consequently more evolved types.
Fig.~\ref{Figure 20} is a histogram showing the numbers of
previously known and new PNe according to their measured image
diameters. Many of the previously known PNe suffer an incremental
increase in diameter due to the luminosity dependent PSF in the
stacked H$\alpha$ maps. The growth function approximates
empirically to
\begin{displaymath}
D = \frac{M_{H\alpha} - 18.557}{-0.3803}
\end{displaymath}
where D is the image diameter in arcsec. This image effect has
been caused by the high luminosity of most of these compact and
young PNe. The majority of the new PNe are larger and fainter by
comparison. Because most of them are $>$4 arcsec diameter, and
faint, the UKST map is able to provide a truer, resolved diameter.

Within our initial 3.25$^{\circ}$ square survey area, the PSF of
the H$\alpha$ emission for known PNe is an average $\sim$2.69
times the diameter of the PNe as measured in the SR images. The
same ratio for our newly confirmed PNe gives an average 2.29 times
the SR diameter indicating the new sample are slightly fainter in
H$\alpha$ compared to SR. Conversely, many PNe, both known and
newly confirmed, can only be faintly detected in the SR image and
some not at all. Within the 2dF sampled area, discussed here, this
applies to 4 of the 31 previously known PNe and 34 of the 73 newly
confirmed PNe. In the colour combined images they appear as small,
discrete blue nebulae (eg. RP442 in Fig.~\ref{Figure 5}). By
comparison, emission line stars have an H$\alpha$ PSF only
$\sim$1.65 times larger than the red image diameter. Visually
therefore, the H$\alpha$ halos for emission line stars are more
compact and fade gradually at the edges. Variable stars appear to
have an uneven mix of SR versus H$\alpha$ emission in their halos.
This is due to the time-averaging process of stacking the images
taken over a three year period. Late-type stars are easy to detect
due to the predominance of a pale mixture of the two colours and
star-like appearance. \label{section 4.1}

 \label{section 4}

\section{Discovery of LMC PN optical halos and their significance}
Perhaps the most exciting development is our detection, for the
first time, of large, extended, optical halos around 60 per cent
of both known and new LMC PNe. This fraction agrees with that
determined by Corradi et al. (2003) from deep imaging of 35
elliptical Galactic PNe. In Fig.~\ref{Figure 3} we present a
Hubble Space Telescope (HST) image of the LMC PN SMP27 which is
the only previous LMC PNe for which a tentative halo detection was
previously suggested (Shaw et al. 2001). Adjacent to this we
present our H$\alpha$/SR combined image reproduced at the same
scale. Our stacked image does not have the resolution of the HST
but the data go as deep for low surface brightness features; the
halo emission is clearly seen, matching that seen by the HST. In
Fig.~\ref{Figure 4} we present a wider angle image of this same PN
showing that the halo hinted at in the HST image in fact extends
almost right around the central PN. Beside this image we also
include a more typical example of a newly discovered LMC PN halo.
RP142 in Fig.~\ref{Figure 6}.


The typical halo radii range from 0.4 to 1.5 parsecs, in good
agreement with that found for Galactic PNe by Corradi et al.
(2003) despite significant distance uncertainties and also in
agreement with the spread of faint halo radii predicted by
Villaver et al. (2002) for a range of initial progenitor masses.
Detection of these LMC PN halos, which essentially represents the
AGB mass loss history, will provide a valuable tool to derive the
initial to final mass ratio for stars $<$8M$_{\odot}$.

\begin{figure}
\begin{flushleft}
  \includegraphics[width=.45\textwidth]{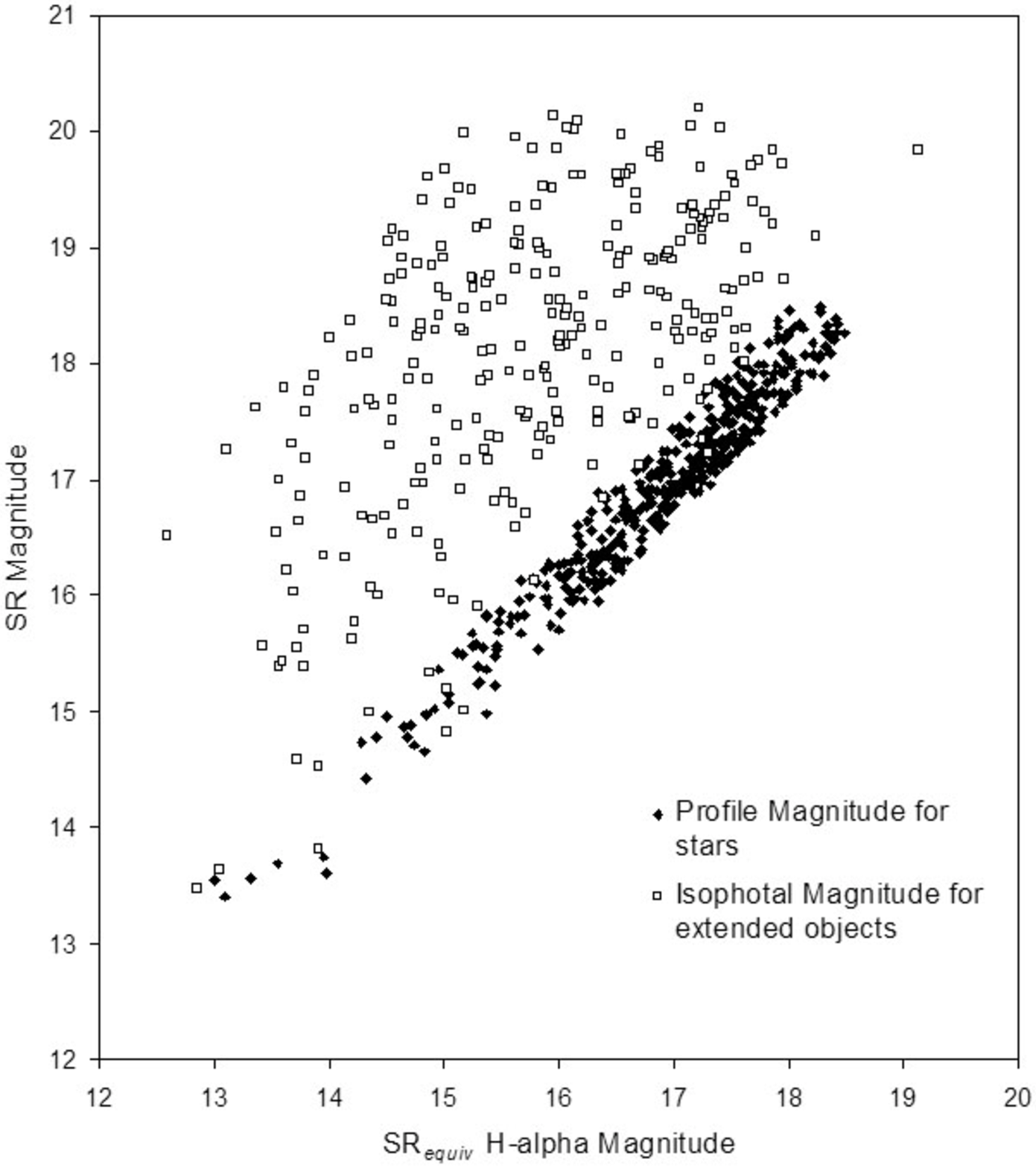}\\
  \caption{SuperCOSMOS R$_{equiv}$ H$\alpha$ and SR magnitudes are compared for stars using PRFMAG and for extended objects using the COSMAG parameter. Stellar sources measured with PRFMAG deviate by $\pm$0.3 magnitudes indicating high consistency of the SuperCOSMOS IAM data. Extended objects measured with COSMAG include PNe, galaxies, emission objects and discrete \HII ~regions. See section~\ref{section 6}.}
  \label{Figure 15}
  \end{flushleft}

\begin{flushleft}
  \includegraphics[width=.45\textwidth]{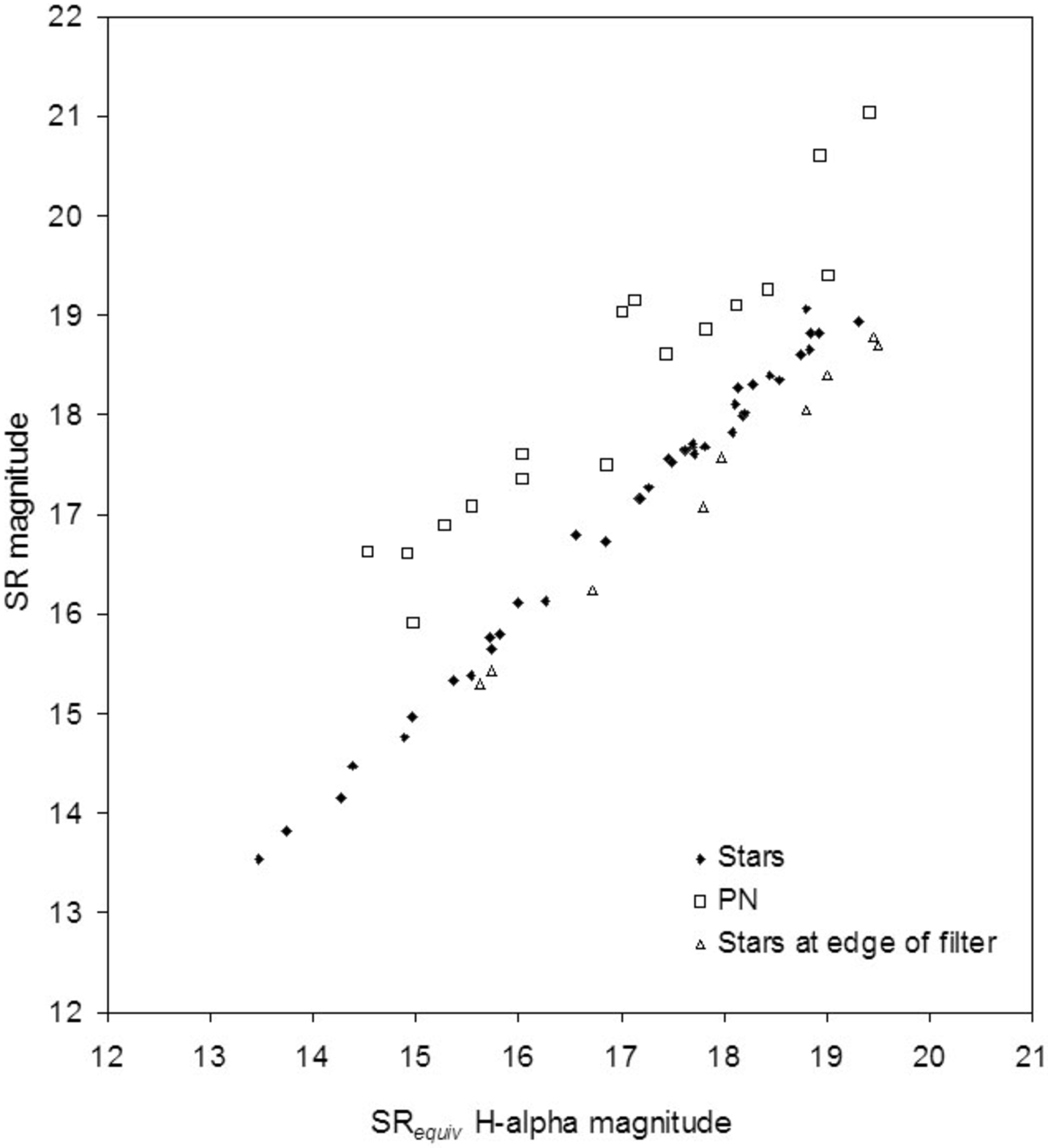}\\
  \caption{R$_{equiv}$ H$\alpha$ versus SR magnitudes for stars, PN and stars at the edge of the H$\alpha$ filter showing the accuracy of manual, de-blended photometry. All magnitudes were calibrated to the SuperCOSMOS IAM magnitudes and derived manually using the {\small PHOTOM} package as described in section~\ref{section 6}.}
  \label{Figure 14}
  \end{flushleft}
  \end{figure}

\section{Photometry}

Along with the pixel data, SuperCOSMOS provides an Image Analysis
Mode (IAM) file which contains 32 image parameters relating to
each object (Hambly et. al. 2001). Within the IAM parameter sets
for each object there are two instrumental magnitudes called
COSMAG and PRFMAG. The COSMAG parameter is an isophotal magnitude
(-2.5 log {sum of intensities of thresholded pixels above sky})
while the PRFMAG parameter is a `linearised' profile magnitude
computed following the prescription of Bunclark \& Irwin (1983).
The photometric calibration for stars is based on profile
magnitudes, while integrated magnitudes for extended sources are
calibrated isophotal magnitudes, sky divided and corrected to
total magnitudes. Using the Starlink {\small GAIA} package, the
IAM parameters for each object can be linked to the appropriate
object in the image which is identified by an over-plotted
ellipse. The ellipse contains the area used to produce the
integrated sky subtracted magnitudes.

SuperCOSMOS magnitudes are calibrated to bright photometry from
the Tycho-2 (Hog, 2000) and Guide Star Photometric Catalogues
(Lasker, 1988), faint CCD standards from Croom et al. (1999),
Boyle et al. (1995), Stobie et al. (1985), Cunow et al. (1997) \&
Maddox et al. (1990).

Although this method has provided roughly calibrated magnitudes
for both the SR and H$\alpha$ images, blended sources became a
problem in the crowded field of the LMC. SuperCOSMOS fails to
effectively de-blend multiple, closely spaced objects. For the
blended PN candidates therefore, we decided to do individual,
manual photometry on each object using {\small GAIA} and the
{\small PHOTOM} package to try to improve the photometry. This
allowed us to carefully place apertures around the exact perimeter
of the object, avoiding the influence of nearby stars and
diminishing the added flux from overlapping stars. This was
achieved by careful calibration of {\small PHOTOM} parameters
until test results on bright and faint stars returned identical
results to the IAM magnitudes for our stacked SR image. We then
used de-blended stars to test our stacked SR magnitudes against
the same stars in our stacked H$\alpha$ data. In the
figure~\ref{Figure 14} test sample, we found that ordinary
continuum stars show very good agreement between their measured SR
and R$_{\textit{equiv}}$ H$\alpha$ magnitudes when manual
de-blended photometry is applied using the {\small PHOTOM}
package. The average displacement between SR and
R$_{\textit{equiv}}$ H$\alpha$ magnitudes across the sample is
0.04 magnitudes with a standard deviation of 0.13 magnitudes. By
comparison, the PN have an average displacement of 1.77 magnitudes
and standard deviation of 0.23 magnitudes due to their brightness
in H$\alpha$.

In the UKST stacked data, positional and magnitude-dependent
systematic errors are present as a result of vignetting towards
the corners of the image (Hambly et al. 2001). IAM stellar
parameters also vary as a function of field position arising from
variable PSF due to field rotation. Other common errors due to
variations in emulsion sensitivity are alleviated by the multiple
stacking process. Re-calibration of {\small PHOTOM} parameters is
therefore required beyond any given area of 1$^\circ$ square. In
our case the situation is further complicated by the use of the
H$\alpha$ interference filter which has a circular aperture of
about 305mm diameter on the square glass substrate (see Parker et
al. 2005). The circular edge of the H$\alpha$ filter came very
close to the 4 corners of the 25$^\circ$ square H$\alpha$ exposure
frame. This resulted in an incremental dimming effect from
$\sim$18$^\prime$ towards the extreme corners on the H$\alpha$ map
making magnitudes within these small zones unreliable. This
effect, plus the above mentioned vignetting and field rotation
effects, have also been measured using various magnitude ranges at
the far edge of the usable H$\alpha$ filter (see fig.~\ref{Figure
14}). We find a maximum dimming effect of $\sim$0.56 of an
H$\alpha$ magnitude with a standard deviation of 0.18 magnitudes
at the edge of the filter. This correction has been applied to the
H$\alpha$ magnitude of any PN found within this region.

A simple test was conducted to measure the accuracy of the
magnitude parameters for both the H$\alpha$ and SR stacked images.
The IAM magnitude parameters for a sample of over 2,000 objects
were extracted from a 25 square arcmin area near the centre of the
LMC field. Matching of the H$\alpha$ and SR IAM object positions
was achieved using the Starlink {\small CURSA} package. Our
selection criteria required object positions in both images to
match to within 0.03 arcsec. Only these objects with perfectly
matching positions were extracted. We then eliminated all blended
objects. This left us with 402 objects identified as stars plus
275 extended objects. The resulting PRFMAG H$\alpha$ and SR
parameter values for the stellar objects were found to be in good
agreement. The average magnitude difference between H$\alpha$ and
SR was 0.028 with a standard deviation of 0.22. The extended
objects, including PNe, galaxies, discrete \HII ~regions and other
emission objects, were plotted using the COSMAG parameter values.
On average, the H$\alpha$ magnitudes for the extended sources were
brighter by 2.34 magnitudes with a standard deviation of 1.14. The
results are plotted in Fig.~\ref{Figure 15}.

\begin{figure}
\begin{flushleft}
  \includegraphics[width=.47\textwidth]{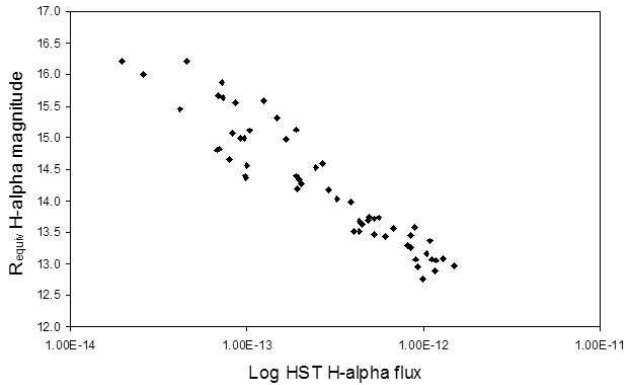}\\
  \caption{UKST R$_{equiv}$ H$\alpha$ magnitudes versus Hubble Space Telescope derived H$\alpha$ fluxes ( in $ergs~cm^{-2}~s^{-1}~$\AA$^{-1}$) for previously known PNe in the LMC.}
  \label{Figure 21}
  \end{flushleft}
  \end{figure}

As an external test, we compared our H$\alpha$, R-equivalent image
derived magnitudes with the MCPN published HST H$\alpha$ fluxes
(http://archive.stsci.edu/hst/mcpn/home.html) for known PNe in the
LMC (Fig.~\ref{Figure 21}). While the comparison of a photographic
image magnitude (which includes the [NII] lines) with a ccd
derived flux centered on H$\alpha$, cannot be expected to match
perfectly, it does produce a reasonably even gradient.

The measured SuperCOSMOS SR magnitudes for the initial sample are
given in tables~\ref{Table 3}~and~\ref{Table 4}. These magnitudes
are the least likely to be affected by the added flux from
overlapping stars. Where de-blending was performed, the image
derived magnitudes have been indicated with a dagger. For only
three objects (SMP72, RP590 and RP491), overlapping stars have
obscured half or more of the visible PN.

Accurate calibrated flux measurements will also be obtained from
our continuing program of follow-up slit spectroscopic
observations over our sample of new PNe.


\label{section 6}

%
%


\section{Confirmatory observations }

\begin{figure}
\begin{flushleft}
  \includegraphics[width=.47\textwidth]{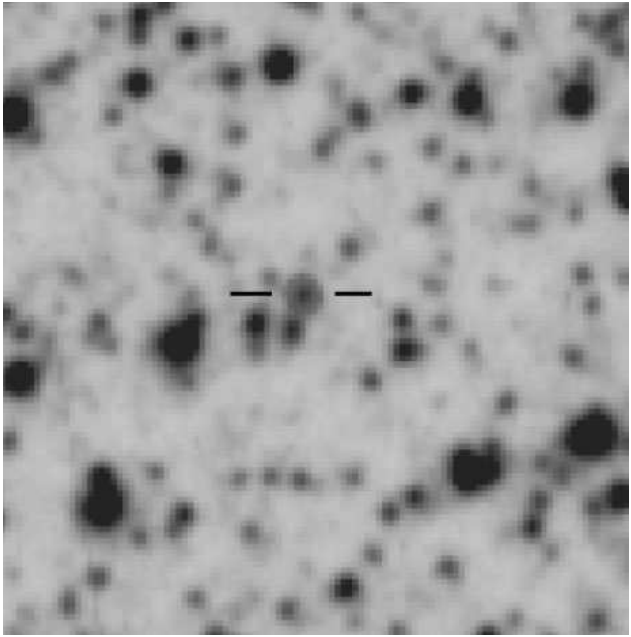}\\
  \caption{Combined H$\alpha$/SR greyscale image of the whole LMC 5$^{\circ}$ $\times$ 5$^{\circ}$ survey area. The current 2dF-observed region is indicated in the bottom left hand corner (cells 1,1, 1,2 and 2,1).}\label{Figure 11}
  \end{flushleft}
  \end{figure}

Initial confirmation of a preliminary selection of LMC PN
candidates was undertaken on 26th November 2003 using 2dF on the
Anglo-Australian Telescope. A short 30 minutes of service time was
sufficient to observe most of the candidates in a 2 degree
diameter area centered on J2000 RA 05 44 46 Dec -70 33 20. 184
objects were observed within this area comprising all of our image
cell `1,1' and half of image cell `1,2'. On the 15th March 2004,
another short service run on the AAT allowed us to observe a
further 81 objects in image cell `2,1', centered at position J2000
RA 05 29 20 Dec -70 52 25. In all, 265 objects have been observed
in the three image cells indicated in Fig.~\ref{Figure 11}. As
well as our new PN candidates we included all known PN,
emission-line stars, late-type stars, variable stars and small
\HII~bubbles in the region for confirmation of their spectral
signatures. We were able to confirm all 31 previously known PN
within the observed image cells and 73 new PN. The remaining
emission sources included emission-line stars, \HII~regions,
late-type stars, variable stars and objects whose status remains
uncertain (due to low S/N). The identification is shown in
Table~\ref{Table 2}. A further 270 emission sources lie in this
area and will be published following spectral analysis. In such a
rich field as the LMC, not every object can be included in one 2dF
observation due to fibre spacing requirements on the field plate.
The number density of the sources also results in fibres being
unable to overlap each other and reach targets. \label{section 7}

\subsection{Position and spectroscopic confirmation of previously known PN}

Of the near 300 PNe previously known to exist in the LMC, 168 fall
within the 25 square degree area of our new UKST deep survey.
Positions for all previously known PNe in our data showed very
good agreement with positions provided by Leisy et al. (1997). We
found that positions provided by earlier surveys (mostly using the
FK4 system) were not sufficiently accurate when converted to the
J2000 equinox. The K-view program in {\small KARMA} allowed us to
find the position of peak intensity within both the stacked
H$\alpha$/SR images and SuperCOSMOS on-line images allowing
accurate positioning to 0.2~arcsec. Where objects have extended
emission, with no visible core or range in flux intensity across
the nebulae, we defined the physical centre of the visible object
as our position reference.

In Table~\ref{Table 3} we present results of our 2dF spectroscopic
observations of the 31 previously known PNe within the 2dF regions
observed in this pilot study. These are presented in order to make
direct [\NII]/H$\alpha$ line ratio comparisons with our newly
detected PNe in the same immediate LMC location. In successive
columns we give the known name, followed by the right ascension
and declination in J2000, the relative size of the central PN seen
as a concentration in the SR once the H$\alpha$ and SR are merged,
PN dimension in the red image, PN dimension in the H$\alpha$
image, an [\NII]6548+6583\AA/H$\alpha$ raw intensity ratio and the
SR magnitude (section~\ref{section 6}). The dimensions are image
dimensions as described in section \ref{section 4.1}. Where one
number is stated, the PN is morphologically round and the
dimension represents the diameter. Where two numbers are stated,
the first refers to the diameter of the long axis and the second
is the diameter of the short axis. Previously known PN, MG79 has
been omitted as we have found no spectral or visual evidence to
indicate the presence of a PN.

\subsection{Spectroscopic confirmation of new candidates}

\begin{table}

\caption{Spectroscopic results from the initial 2dF observations
covering an area $\sim$3.25$^{\circ}$ square, ~SW of the main LMC
bar; seen in Fig.~\ref{Figure 11} (270 further candidates in this
area are awaiting spectroscopic ID).}
\begin{center}
\begin{tabular}{|l|c|cl}
  \hline
  Object &  Previously &  Newly \\
  & Known &  Confirmed  \\
  \hline
  PN & 31 & 73 \\
  Emission-line stars & 14 & 36\\
  Late-type stars  & 8 & 44  \\
  \HII~ regions  & 7 &   \\
  Other stars   & 2 &  5  \\
  Variable stars  & 5  & 2  \\
  Emission objects of &   & \\
  unknown nature & 2 & 6  \\
  SNR     & &  1 \\
  S/N too low for ID  & &  29  \\
  \hline \noalign{\smallskip} 
  \label{Table 2}
  \end{tabular}
\end{center}
\end{table}

In Table~\ref{Table 4} we present our confirmed LMC PN discoveries
by their RP catalogue name and newly assigned IAU nomenclature.
Columns 3 and 4 list the RA and Dec in J2000, while column 5
presents the relative size of the central PN (excluding the halo)
in terms of its central concentration once the SR and H$\alpha$
images have been merged. Objects referred to as `H$\alpha$ only'
in this column are not visible at all in the SR band once the
H$\alpha$ image has been merged (see section~\ref{section 4} and
the example image of RP442 in Fig.~\ref{Figure 5}). Columns 6 \& 7
present the dimension in arcsec as measured separately in the SR
and H$\alpha$ images. The [\NII]6548+6583\AA/H$\alpha$ raw
intensity ratio from the 2dF spectra and the SR magnitude
(section~\ref{section 6}) are presented in columns 8 \& 9.

Candidates were selected as PNe by the combined examination of
image parameters and spectra. In the spectra we examined the main
ratios: $\textit{I}~$\OIII 5007\AA/H$\beta$, $\textit{I}~ $\OIII
5007\AA/H$\alpha$ and $\textit{I}~ $[\NII] 6583\AA/H$\alpha$. We
also measured the strength of the [SII](6716/6731\AA) lines and
took the presence of HeII~4686 into account as this indicates a
high level of excitation, very rarely seen in \HII~regions. Every
other emission line present was also carefully measured. Allowance
was made for a degree of continuum in the spectrum where we could
see PNe partially hidden by intervening stars (2dF fibres are 2.5
arcsec in diameter).

\begin{figure}
\begin{flushleft}
  \includegraphics[width=.46\textwidth]{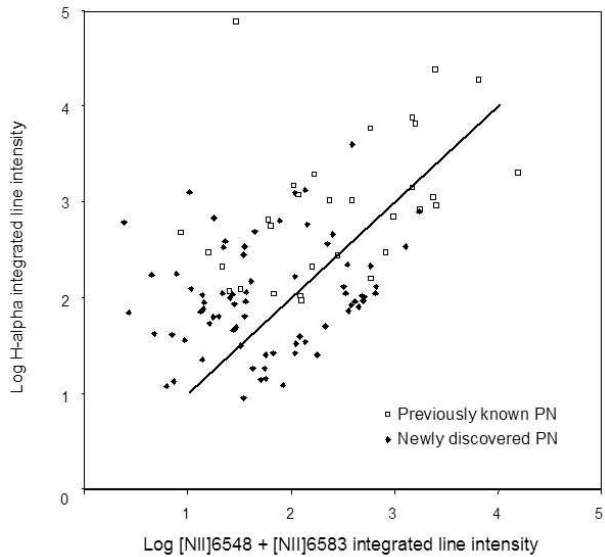}\\
  \caption{Plot of Log H$\alpha$ versus Log [NII] 6548 + 6583 relative raw line intensities for previously known and newly confirmed PN in the observed region.}\label{Figure 13}
  \end{flushleft}
  \end{figure}

Fig.~\ref{Figure 13} shows an initial log plot of the [\NII]~6548
+ 6583\AA ~versus H$\alpha$ raw, integrated, relative line
intensities. A full description of ratios and excitation classes
from the new sample across the entire LMC 25$^{\circ}$ square
survey area will be presented in subsequent papers following
completion of the spectroscopic followup and photo-ionization
modelling. \label{subsection 7.2}

\section{The  new LMC emission object catalogue}

We have created a relational database to store and collate all the
gathered LMC data for this project. This facilitates the
collection of information such as probability ratings, diameters,
magnitudes, chemical abundances and line ratios from our
confirmatory spectroscopy. We can then create queries and reports
based on any aspect of the stored data. Reports are created on
each individual object which include a 50 square arcsec combined
colour relief image and spectrum for instant confirmation. All
previously known PN in the LMC have been included along with all
their known names, magnitudes, fluxes and velocities.

In addition to the new and previously known PNe, the final
catalogue will comprise over 2,000 emission sources including both
known and newly discovered emission-line stars, WR stars,
late-type stars, SNRs, small (bubble-type) \HII~regions, variable
stars and other emission objects. Accurate J2000 coordinates will
accompany each object. The whole LMC PN catalogue, will be
published and available on-line within the next 12 months complete
with spectra and images for each source.

We are currently in the process of finishing a detailed
spectroscopic followup of all remaining candidates. Results from
these observations will be released through further papers in this
series.

\begin{table*}

\caption{Previously known PN within the spectroscopically
confirmed fields}
\begin{tabular}{|l|c|c|l|c|c|c|c|c|}
  \hline
  Catalogue Name*  & RA & Dec & Central PN & SR dim. & H$\alpha$ dim.
  &  [\NII]/H$\alpha$ & SR mag
  \\
  &   (J2000)  & (J2000)  &   observed  &  (arcsec)  &  (arcsec)  &
  intensity  &   \\
  \hline
SMP55, LM2-21, N199, LI924 &       05 22 40.96 &   -71 19 06.7  &  Medium &   7    &   11.4$\times$10.4    &  0.23    &   15.5   \\
Mo24  &            05 22 53.20 &   -71 05 40.7  &  Point  &   4 &  5.3 &   1.19  &   17.8       \\
SMP60 &            05 24 15.69 &   -70 53 56.3  &  Point &   6    &   8$\times$6.6 &  0.07   &   16.5     \\
SMP59, LM2-25 &    05 24 27.35 &   -70 22 23.7  &  Small &   8 &   10 &    2.83   &   16.1    \\
MG43 &             05 24 34.29 &   -71 13 39.6  &  H$\alpha$ only &     &   5.3     & 0.05   &     \\
SMP62, WS25, LM1-38, N201 &     05 24 55.04 &   -71 32 55.4  &  Large &   11.3$\times$9.1    &   14$\times$12  &   0.10    &   14.6      \\
SMP65 &            05 27 43.83 &   -71 25 56.0  &  Point &   5.3$\times$5   &   7.3$\times$7    &  0.02   &   16.7     \\
MG51  &            05 28 34.41 &   -70 33 01.6  &  Point &   4.6 &   6   &       &   17.3$\dag$          \\
SMP68 &            05 29 02.85 &   -70 19 24.8  &  Small &   4.6$\times$4    &   8.7$\times$8 &   0.01  &   17.1      \\
Mo28 &             05 29 18.43 &   -70 23 49.7  &  Point  &   4 &   5.3   &   1.04   &   17.4$\dag$       \\
Sa120 &            05 29 32.70 &   -70 17 39.0  &  Point &   4    &   8$\times$7   &   2.15   &   17.5       \\
Sa121 &           05 30 26.27 &   -71 13 48.0  &  Point  &   4    &   6   &   1.07   &   17.8        \\
SMP71, N207 &           05 30 33.30 &   -70 44 37.6  &  Medium &   4.7    &   11  &   0.35    &   15.2        \\
SMP72 &           05 30 45.83 &   -70 50 15.8  &  H$\alpha$ only &    &   7.3$\times$6.4   &    0.08   &          \\
SMP73, N208 &           05 31 21.97 &   -70 40 44.9  &  Medium &   8$\times$7    &   15$\times$13.2  &   0.2    &   14.6         \\
Mo33  &           05 32 09.28 &   -70 24 41.5  &  H$\alpha$ only &   &   7.3    &   0.27  &  \\
Sa123, LM2-36 &   05 34 30.17 &   -70 28 34.5  &  Small &   5.7   &   10 &  2.18   &   16.4      \\
SMP80, LM2-37&    05 34 38.95 &   -70 19 55.5  &  Medium &   4.6    &   9.4   &   0.1    &   15.9       \\
Mo35 &            05 38 04.62 &   -70 29 25.9  &  Small  &   4 &   8$\times$7   &  3.78   &   16.5      \\
Mo36 &            05 38 53.56 &   -69 57 55.7  &  Small  &   6$\times$5.2 &   8$\times$7   &  1.38    &   17.1    \\
Mo37 &             05 39 14.47 &   -70 00 18.6  &  Point  &   4 &   6$\times$5   &   0.10   &   18.3     \\
LM2-40  &          05 40 37.60  &  -70 27 54.97  &  Point  &  4.7 & 8  &  0.21 &  17.7$\dag$ \\
MG73  &            05 41 36.61 &   -69 27 09.8  &  Small  &   4.7 &   6   &   2.85    &   17.3$\dag$     \\
MG76 &             05 42 24.24 &   -69 53 05.1  &  Medium  &   4    &   5.3    &  0.11   &   16.9$\dag$      \\
SMP88, LM2-42 &    05 42 33.24 &   -70 29 23.2  &  Large  &   7.9$\times$7.3 &   8.4$\times$8 &   0.37    &   15.4    \\
SMP89, LM156, N178, WS38  &   05 42 37.00 &   -70 09 31.1  &  Large  &   7.4    &   10  &  0.1    &   14.4$\dag$     \\
Mo39 &             05 42 41.09 &   -70 05 49.1  &  H$\alpha$ only  &    &   4   &   0.76   &       \\
SMP90, LM2-43 &    05 44 34.76 &   -70 21 40.5  &  Medium  &   6.3$\times$4 &   7.4$\times$6.8 &  1.39   &   16.9    \\
Mo40 &            05 46 25.29 &   -71 23 22.3  &  Point  &   4 &  4  &   0.62   &   18.5       \\
SMP92, N170, LM1-58, WS39 &            05 47 04.37 &   -69 27 32.9  &  Large  &   9 &   13.6   &   0.2   &   14.1     \\
SMP93, N181 &           05 49 38.75 &   -69 09 59.3  &  Large  &   10.8$\times$6.6    &   13.9$\times$10   &  7.92   &   15.0    \\
\hline \noalign{\smallskip} 
\label{Table 3}
\end{tabular}

* Explanation of abbreviations used: ~LI: Lindsay (1963),
~LM1: Lindsay \& Mullan (1963), ~LM2: Lindsay \& Mullan (1963),
~MG: Morgan \& Good (1992), ~Mo: Morgan (1994), ~N: Henize (1956),
~SMP : Sanduleak et
al. (1978), ~Sa: Sanduleak (1984), ~WS: Westerlund \& Smith (1964) \\



\caption{Newly discovered PN within the spectroscopically
confirmed fields}
\begin{tabular}{|l|c|c|c|l|c|c|c|cl}
  \hline
  Cat. Name & IAU Name & RA & Dec & Central PN & SR dim. & H$\alpha$ dim.
  &   [\NII]/H$\alpha$ & SR mag
  \\
  &  &   (J2000)  & (J2000)  &   observed  &  (arcsec)  &  (arcsec)  &
  intensity  &   \\
  \hline
RP590 & RPJ 052223-703355   &     05 22 23.98 &   -70 33 55.9    &   H$\alpha$ only   &      &   4   &  0.61 &      \\
RP478 & RPJ 052612-705855   &     05 26 12.92 &   -70 58 55.4    &   Medium  &   5.5$\times$4   &   7$\times$6   & 3.71 &   16.3      \\
RP589 & RPJ 052637-702907   &     05 26 37.58 &   -70 29 07.1    &   H$\alpha$ only   &     &  5.3   & 0.25 &    \\
RP561 & RPJ 052639-710027   &     05 26 39.31  &   -71 00 27.57  &     H$\alpha$ only  &     &  4  &  0.66 &   \\
RP594 & RPJ 052711-702623   &     05 27 11.03 &   -70 26 23.3    &   H$\alpha$ only &   &   4  & 0.55  &      \\
RP607 & RPJ 052803-704239   &     05 28 03.28 &   -70 42 39.6    &   H$\alpha$ only &   & 4   &  2.72 &    \\
RP523 & RPJ 053005-711347   &     05 30 05.17 &   -71 13 47.9    &   H$\alpha$ only &  &  5.3    & 0.33  &   \\
RP427 & RPJ 053010-704639   &     05 30 10.87 &   -70 46 39.6    &   H$\alpha$ only  &       &   7.3$\times$6.7  &  4.69 &      \\
RP415 & RPJ 053157-704646   &     05 31 57.97 &   -70 46 46.2    &   Point    &   4    &   4.7  & 0.02  &   19.9     \\
RP525 & RPJ 053229-711101   &     05 32 29.67 &   -71 11 01.8    &   H$\alpha$ only &  &     7.6$\times$6   &  3.93   &    \\
RP441 & RPJ 053242-703840   &     05 32 42.82 &   -70 38 40.5    &   Small   &   4 &   7.3   & 3.37 &   18.0      \\
RP442 & RPJ 053251-703717   &     05 32 51.35 &   -70 37 17.0    &   H$\alpha$ only &   &   4   & 0.19  &    \\
RP548 & RPJ 053308-711803   &     05 33 08.75 &   -71 18 03.3    &   Small  &  7  &  4.7  &  &  17.7  \\
RP604 & RPJ 053328-712450   &     05 33 28.13 &   -71 24 50.7    &   H$\alpha$ only &  &  4.7    & 0.28  &   \\
RP603 & RPJ 053329-712451   &    05 33 29.62 &   -71 24 51.0    &   H$\alpha$ only &   &   7 & 0.12 &    \\
RP499 & RPJ 053459-710606   &    05 34 59.44 &   -71 06 06.8    &   Point  &  4  &  6  &  1.59  &  18.7  \\
RP530 & RPJ 053548-710627   &    05 35 48.68 &   -71 06 27.1    &   Small  &  4  &  7.3   & 0.02  &  18.8 \\
RP491 & RPJ 053611-711719   &    05 36 11.39 &   -71 17 19.1    &   H$\alpha$ only   &    &  7.3   & 0.01  &   \\
RP577 & RPJ 053632-702925   &    05 36 32.84 &   -70 29 25.7    &   H$\alpha$ only  &   &7.3   & 0.60  &    \\
RP232 & RPJ 053635-692228   &    05 36 35.22 &   -69 22 28.7    &   Medium  &   4 &   6     & 0.17 &   18.0     \\
\multicolumn{8}{r}{continued next page $\rightarrow$}\\
\hline
\label{Table 4}
\end{tabular}
\end{table*}

\begin{table*}
\begin{tabular}{|l|c|c|c|l|c|c|c|cl}
&{\it (cont'd)}\\
  \hline
  Cat. Name & IAU Name  & RA & Dec & Central PN & SR dim. & H$\alpha$ dim.
  &   [\NII]/H$\alpha$ & SR mag
  \\
  &  &   (J2000)  & (J2000)  &   observed  &  (arcsec)  &  (arcsec)  &
  intensity  &   \\
  \hline
RP580 & RPJ 053638-702505   &    05 36 38.29 &   -70 25 05.5    &   Point    &   4 &   6   & 1.4 &   18.6      \\
RP234 & RPJ 053641-692208   &    05 36 41.16 &   -69 22 08.3    &   Point    &   4$\times$3.4 &   6$\times$4     & 0.13 &   18.2      \\
RP231 & RPJ 053649-692355   &    05 36 49.38 &   -69 23 55.2    &   Point    &   4.7 &   4     & 0.05 &   16.5    \\
RP265 & RPJ 053700-692128   &    05 37 00.79 &   -69 21 28.3    &   Point    &   4 &   5.4$\times$4.5   & 2.16 &   16.4    \\
RP228 & RPJ 053706-692709   &    05 37 06.65 &   -69 27 09.3    &   Point   &   4  &  4   &  0.04  &   17.9  \\
RP26  & RPJ 053707-701951   &    05 37 07.46 &   -70 19 51.0    &   H$\alpha$ only  &   &   4$\times$3.5     & 0.11 &        \\
RP493 & RPJ 053710-712313   &    05 37 10.14 &   -71 23 13.9    &   Small  &  4  &  4     &  0.09  &  17.5  \\
RP1040 & RPJ 053721-700408  &    05 37 21.05  &  -70 04 08.40   &    H$\alpha$ only  &  &  4  &  0.12  &   \\
RP1037 & RPJ 053725-694759  &    05 37 25.26  &  -69 47 59.82   &   H$\alpha$ only &   &  4  &  0.56  &  \\
RP266 & RPJ 053727-690855   &    05 37 27.77 &   -69 08 55.2    &   Small   &   5   &   9.6   & 0.09 &   15.3     \\
RP147 & RPJ 053729-700750   &    05 37 29.28 &   -70 07 50.9    &   small  &   4.6 &   14    & 2.21 &   17.7$\dag$  \\
RP25  & RPJ 053729-701633   &    05 37 29.44 &   -70 16 33.0    &   large  &   5 &   6    &  3.1 &   16.6$\dag$      \\
RP87  & RPJ 053810-703245   &    05 38 10.55 &   -70 32 45.5    &   H$\alpha$ only   &    &  7    &   6.92  &   \\
RP144 & RPJ 053838-701325   &    05 38 38.28 &   -70 13 25.5    &   H$\alpha$ only  &    &   5$\times$4     & 3.99 &     \\
RP90  & RPJ 053840-701901   &    05 38 40.85 &   -70 19 01.4    &   Small   &   4 &   4.6$\times$4    & 0.30 &   18.4   \\
RP89  & RPJ 053856-702121   &    05 38 56.41 &   -70 21 21.6    &   Point    &   6$\times$4.6   &   6$\times$4.6    & 0.25 &   18.6    \\
RP182 & RPJ 053905-695045   &     05 39 05.12 &   -69 50 45.9    &   H$\alpha$ only &      &   7     & 1.03 &      \\
RP143 & RPJ 053931-700615   &    05 39 31.25 &   -70 06 15.4    &   Point    &   4.7$\times$3.6 &   11$\times$9.3     & 4.56 &   17.2$\dag$     \\
RP35  & RPJ 053950-712801   &    05 39 50.73 &   -71 28 01.2    &   Point    &   4 &   7.2$\times$6   &  4.3&   19.1     \\
RP44  & RPJ 053956-710922   &    05 39 56.44 &   -71 09 22.6    &   Point    &   4    &   5.3$\times$4.6     & 0.54 &   18.6     \\
RP241 & RPJ 054020-691300   &    05 40 20.68 &   -69 13 00.8    &   Point    &   4.7 &   6    & 0.06 &   17.7    \\
RP178 & RPJ 054028-695439   &    05 40 28.51 &   -69 54 39.5    &   Point  &   4 &   4    & 0.16 & 19.3$\dag$     \\
RP105 & RPJ 054045-702806   &    05 40 45.26 &   -70 28 06.7    &   Large   &   11$\times$8 &   12.6$\times$9.3    & 0.10 &   14.0$\dag$      \\
RP125 & RPJ 054053-704508   &     05 40 53.51 &   -70 45 08.2    &   Point    &   4  &  6     & 3.91  &   19.8$\dag$  \\
RP111 & RPJ 054127-703207   &     05 41 27.52 &   -70 32 07.9    &   H$\alpha$ only  &    &   5    & 2.48 &     \\
RP397 & RPJ 054152-702818   &    05 41 52.95 &   -70 28 18.8    &   H$\alpha$ only &  &  5.3$\times$4   &  7.09 &    \\
RP93  & RPJ 054202-702459   &    05 42 02.16 &   -70 24 59.7    &   Point    &   4 &   5    & 5.94 &   17.3      \\
RP61  & RPJ 054226-704940   &    05 42 26.21 &   -70 49 40.4    &   Point    &   4.6 &   6.5  & 4.64 &   17.5     \\
RP187 & RPJ 054236-694023   &    05 42 36.06 &   -69 40 23.6    &   H$\alpha$ only  &   &   7   & 0.32  &    \\
RP9   & RPJ 054236-702859   &    05 42 36.49 &   -70 28 59.3    &   H$\alpha$ only &    &   5.7   & 0.60 &      \\
RP135 & RPJ 054238-700438   &    05 42 38.83 &   -70 04 38.4    &   Small   &   4   &   4    & 5.32 &   19.4  \\
RP70  & RPJ 054302-703332   &    05 43 02.15 &   -70 33 32.2    &   H$\alpha$ only  &  &  8    & 3.7  &  \\
RP162 & RPJ 054317-695651   &    05 43 17.63 &   -69 56 51.4    &   H$\alpha$ only  &    &   6.1     & 0.522  &       \\
RP621 & RPJ 054346-705804   &    05 43 46.99 &   -70 58 04.1    &   H$\alpha$ only  &   &  8.7$\times$7    &  0.32 &  \\
RP62  & RPJ 054422-704043   &    05 44 22.66 &   -70 40 43.4    &   Point    &   4    &   6    & 0.17 &   21.0     \\
RP163 & RPJ 054428-695443   &    05 44 28.71 &   -69 54 43.6    &   H$\alpha$ only  &     &   6.7    & 0.18 &     \\
RP637 & RPJ 054437-691947   &    05 44 37.27 &   -69 19 47.5    &   H$\alpha$ only & &  5.3    &  0.4  &  \\
RP615 & RPJ 054519-711603   &     05 45 19.22 &   -71 16 03.2    &   Medium &    5.2 & 7.3   & 0.24 &  15.4$\dag$  \\
RP77  &  RPJ 054534-704303   &     05 45 34.15  &  -70 43 03.93  &  Small  &  4  &  10  &  4.4  &  19.3  \\
RP620 & RPJ 054553-705516   &    05 45 53.27 &   -70 55 16.0    &   H$\alpha$ only  &   &   6   & 5.65  &   \\
RP1038 & RPJ 054606-700457   &    05 46 06.92 &   -70 04 57.5    &   H$\alpha$ only  &    &    4.5  &  4.14  &      \\
RP194 & RPJ 054621-693531   &    05 46 21.16 &   -69 35 31.6    &   H$\alpha$ only  &   4  &   4   & 2.98 &   18.6    \\
RP99  & RPJ 054623-702555   &    05 46 23.95 &   -70 25 55.8    &   Point    &   4    &   5   & 0.05 &   19.4     \\
RP1955 & RPJ 054632-705534   &    05 46 32.63 &   -70 55 34.0    &   H$\alpha$ only  &  4 &   5.3   &  5.05 &  20.8  \\
RP1956 & RPJ 054639-705650   &   05 46 39.38 &   -70 56 50.06     &   Point &    4  &  5  &  0.15  &  17.4$\dag$  \\
RP172 & RPJ 054653-695133   &     05 46 53.09 &   -69 51 33.4    &   H$\alpha$ only  &  4   &   4    & 3.08 &   20.4    \\
RP103 & RPJ 054701-702842   &    05 47 01.53 &   -70 28 42.5    &   Point    &   4.6 &   6$\times$5    & 5.166 &   17.2     \\
RP122 & RPJ 054728-702015   &    05 47 28.70 &   -70 20 15.9    &   Point    &    4  &   9    &   2.35     &  18.7$\dag$ \\
RP130 & RPJ 054819-700523   &    05 48 19.38 &   -70 05 23.8    &   Medium  &   5.2    &   6  & 0.18 &   17.8    \\
RP129 & RPJ 054829-700846   &    05 48 29.12 &   -70 08 46.5    &   Medium  &   9    &   12  & 0.08 &   14.2    \\
RP102 & RPJ 054920-702809   &    05 49 20.52 &   -70 28 09.3    &   Small   &   4 &   6   & 5.51  &   18.9    \\
RP18  & RPJ 054953-700855   &    05 49 53.13 &   -70 08 55.6    &   Point    &   4  & 4.7   &  0.13  &   19.1  \\
RP1957 & RPJ 055004-704035   &    05 50 04.33 &   -70 40 35.44    &   H$\alpha$ only  &  & 4.6$\times$4    & 4.9  & \\
\hline
\end{tabular}

$\dag$ PNe magnitudes for which manual de-blended image
photometry was performed. \\

\end{table*}



\section{conclusion}

We present initial results from the spectroscopic followup of LMC
PN candidates in a ~3.25$^{\circ}$ square area resulting in the
confirmation of 73 new LMC PNe. Based on the preliminary results
of this study, we expect to triple the number of LMC PNe. The
availability of the stacked images has allowed us to observe $>$1
mag deeper than was previously available for the LMC. This has
resulted in the discovery of bright and faint, compact and
extended and young and evolved PNe. They can now be mined for
their chemical abundances and line ratios. They will assist in
kinematical studies and chemical abundance estimates for the LMC.
They will also allow us to derive an accurate luminosity function
for PNe in this galaxy which can be related to other Hubble-type
galaxies. The ratio of [\NII]$>$H$\alpha$ (55 and 75 per cent)
between the previously known and the new sample respectively shows
that the new sample will be able to extend PNe with strong [\NII]
ratios to fainter magnitudes. The discovery of extended AGB halos
surrounding a large proportion of the surveyed PN will provide a
valuable distance unbiased tool to assist in determining the
initial to final mass ratio for stars $<$8M$_{\odot}$. All these
factors make the LMC PN population a unique and ideal sample for
study.

\section{acknowledgements}

WR acknowledges Macquarie University for a PhD. scholarship to
enable this research. We also thank the AAO for service time with
2dF on the AAT. We wish to thank Letizia Stanghellini and Richard
Shaw for their kind permission to use the HST image. Finally we
thank George Jacoby for his help and input during the review process
for this paper.

\label{lastpage}

\end{document}